# Host-to-host variation of ecological interactions in polymicrobial infections

Sayak Mukherjee[1,2], Kristin E. Weimer[6], Sang-Cheol Seok[1], Will C. Ray[1,2], C. Jayaprakash[1,3], Veronica J. Vieland[1,2,4], W. Edward Swords[6], Jayajit Das[1,2,3,5*]

[1]Battelle Center for Mathematical Medicine, The Research Institute at the Nationwide Children's Hospital and Departments of [2]Pediatrics, [3]Physics, [4]Statistics, [5]Biophysics Graduate Program, The Ohio State University, 700 Children's Drive, Columbus, OH 43205.

[6]Department of Microbiology and Immunology, Wake Forest School of Medicine, Winston-Salem, NC 27101

## Abstract

Host-to-host variability with respect to interactions between microorganisms and multicellular hosts are commonly observed in infection and in homeostasis. However, the majority of mechanistic models used in analyzing host-microorganism relationships, as well as most of the ecological theories proposed to explain co-evolution of host and microbes, are based on averages across a host population. By assuming that observed variations are random and independent, these models overlook the role of inter-host differences. Here we analyze mechanisms underlying host-to-host variations, using the well-characterized experimental infection model of polymicrobial otitis media (OM) in chinchillas, in combination with population dynamic models and a Maximum Entropy (MaxEnt) based inference scheme. We find that the nature of the interactions among bacterial species critically regulates host-to-host variations of these interactions. Surprisingly, seemingly unrelated phenomena, such as the efficiency of individual bacterial species in utilizing nutrients for growth and the microbe-specific host immune response, can become interdependent in a host population. The latter finding suggests a potential mechanism that could lead to selection of specific strains of bacterial species during the coevolution of the host immune response and the bacterial species.

**Introduction:**

Consequences of a pathogen exposure or diversity of resident microbiota often vary from individual to individual in a population. This becomes evident when only one of two colleagues sharing the same office falls sick to a flu outbreak, or in experiments studying infection by specific pathogens in animals kept in controlled facilities, where bacterial or viral titers as well as abundances of biomarkers associated with the host immune response display wide ranges of variation between animals[1-3]. Similar variations between individuals are also observed in the structure of the community of microbes residing in homeostasis with the immune system[4].

However, despite the ubiquity of such host-to-host variations of the host-microorganism relationship, our mechanistic understanding of such relationships or ecology of host-microorganisms[5,6] are based primarily on average values obtained from experiments done on a host population. The variations around the averages are usually assumed to arise due to independent inter-host variations of phenomena that affect the host-microorganism relationship, such as the host immune response or availability of nutrients for the microorganisms, and, variations between hosts are often represented merely as error bars in data summaries[7,8]. But this overlooks the fact that the differences between hosts themselves may provide valuable clues regarding perturbations of the underlying mechanistic framework in a natural setting, and may relate directly to evolutionary selection of a particular host-pathogen or host-microbiota relationship based on sustaining the observed diversity in a population[9].

Here we seek mechanistic insights into host-to-host variations of the host-microorganism relationship by using the well characterized model of polymicrobial otitis media (OM) in adult chinchillas (*Chinchilla lanigera*). OM is a common childhood polymicrobial infection of the middle ear involving one or more of three predominant bacterial species that are normally carried within the microbiota in the upper respiratory tract (URT)[10,11]: Nontypeable *Haemophilus influenzae* (NTHI), *Streptococcus pneumoniae* (Sp), and *Moraxella catarrhalis* (Mcat). OM provides an excellent model system to dissect host-microbiota relationships because of the relatively small number of species in the relevant microbial community, and also because it offers practical advantages such as culturability of the three main bacterial species[11]. While chinchillas are not a natural host for the bacteria or viruses that cause human OM, they can be infected and/or colonized with all three of the predominant bacterial OM pathogens[11].

Using an in silico approach based on Maximum Entropy (MaxEnt) and population dynamics, combined with samples recovered from the chinchilla middle ear, we quantified ecological interactions that regulate kinetics of bacterial infection and the host immune response in individual hosts. We show here that the nature of interspecies interactions (e.g., competition, co-operation or neutral) between the bacterial species NTHI and Sp, which is not directly related to the immune response, critically regulates the host-to-host variations of the ecological interactions. More importantly, seemingly independent ecological interactions, such as the ability of the bacterial species to utilize resources and the rate at which the host immune response eliminates specific bacterial species, become inter-dependent in hosts. This suggests evolutionary selection of interspecies interactions in microbial communities through host-bacteria interactions.

**Variations of kinetics of polymicrobial infection**

Animal-to-animal variations of kinetics of bacterial species are clearly observed in experiments studying OM in rodents such as rats[1] or chinchillas[2]. For instance, in the experiments reported by Weimer et al.[2], the population of Sp showed an almost bimodal behavior at three days post inoculation with mixed NTHI and Sp strains; the Sp population fell below the detectable limit in a few animals, but varied between $10^4$ to $10^6$ CFUs in other animals (Fig. 1). The population kinetics of NTHI, although less dramatic, showed animal-to-animal variations up to three orders of magnitude in the experiments with single and mixed species inoculations (Fig.1). The bacterial species, NTHI and Sp, have been observed to interact with each other and with the host and these interactions affect the growth of the bacterial species. For example, in in vitro cultures certain strains of Sp eliminate NTHI by secreting the toxin hydrogen peroxide generated during aerobic metabolism[12], or, NTHI can trigger mobilization of neutrophils in the epithelial layer that eliminate Sp but not NTHI via complement-mediated opsonization[13,14]. In addition, the secretion of quorum sensing molecules by these bacterial species has been found to affect the growth of multiple bacterial species participating in the infection[15]. The bacterial species also depend on the host for extracting essential nutrients such as metals for their growth. E.g., the Gram-negative NTHI and Gram-positive Sp require iron extracted from the serum generated by the host during inflammation[16,17]. Therefore, it is plausible that variations of these factors across hosts would lead to differences in infection kinetics between hosts. Here we quantify ecological interactions in the system and model the mechanisms that lead to the infection kinetics observed in the experiments reported by Weimer et al. [2].

**MaxEnt based method to quantify variations of ecological niches**

*A. Population dynamic model:* We constructed ordinary differential equation (ODE) based kinetic models to describe the time evolution of populations of NTHI and Sp bacterial cells (Fig. S1). The equations are based on Lotka-Volterra (LV) type models[7], which describe the growth of two or more bacterial species interacting with each other to access available resources. These models have been successfully applied to characterize kinetics of bacterial populations in chemostat experiments[7,18]. We modified the LV models to include the host immune responses during the acute infection phase, which is primarily regulated by innate immunity[11]. In our models, the bacterial species consume nutrients from the local environment and replicate. NTHI and Sp can compete for a common nutrient (e.g., iron) for their growth, and additionally each species can indirectly help in the growth of the other species by generating more inflammation. In addition, NTHI and Sp can affect each other's growth by secreting small molecules, e.g., toxins or quorum sensing molecules. Therefore, NTHI and Sp can potentially oppose, help, or remain uninvolved in each other's growth depending on the nature of inflammation or the concentration of secreted molecules in the microenvironment. We considered all 9 possibilities (see Table I) for inter-species interactions affecting the growth rates of NTHI and Sp.

In addition, both species induce innate immune responses (antimicrobial proteins[19] or influx of neutrophils in epithelial layer[14]) in the middle ear. In our models, we do not distinguish between antimicrobial proteins or neutrophils, and immune response is represented by a single variable, I, that eliminates NTHI and Sp with different rates. The dynamics of the abundances of NTHI and Sp in the presence of the host immune response in the middle ear of a particular animal (indexed by *a*) can be described by a pair of coupled ODEs:

$$dN_{1,2}^{(a)}/dt=f_{1,2}^{(a)}(N_1^{(a)},N_2^{(a)})-g_{1,2}^{(a)}(N_1^{(a)},N_2^{(a)}) \quad (1)$$

where, $N_1^{(a)}$ and $N_2^{(a)}$ denote the population sizes of NTHI and Sp, respectively. $f_1^{(a)}(N_1^{(a)},N_2^{(a)})$ and $f_2^{(a)}(N_1^{(a)},N_2^{(a)})$ describe the growth rate of NTHI and Sp, respectively, regulated by available resources and inter/intra species interactions. Both NTHI and Sp interact with the immune response elicited by the host that eliminates the bacteria, and $g_1^{(a)}(N_1^{(a)},N_2^{(a)})$ and $g_2^{(a)}(N_1^{(a)},N_2^{(a)})$ describe the rate of elimination of NTHI and Sp, respectively, by the immune response. To keep the notation simple, we will drop the superscript in the rest of the equations where all the variables and the parameters describe the kinetics in an individual animal or a trial in culture experiments. Following the LV model for interspecies interaction we use[7], $f_1(N_1, N_2)=r_1 N_1(K_1-\alpha_{11}N_1-\alpha_{12}N_2)$ and $f_2(N_1, N_2)=r_2N_2(K_2-\alpha_{22}N_2-\alpha_{21}N_1)$. The carrying capacities, $\{K_1, K_2\}$, determine the maximum values of the population that can be sustained by the available resources[7]. $\{\alpha_{11}, \alpha_{22}\}$ denote the competition for resources between the bacterial cells in the same species and $\{\alpha_{12}, \alpha_{21}\}$ parametrize interspecies interaction between NTHI and Sp. We have used $\{\alpha_{11}, \alpha_{22}> 0\}$, implying that the bacteria in the same species always compete with each other for resources. We considered positive, negative, and zero values for $\{\alpha_{12}, \alpha_{21}\}$ to describe competing, co-operating, and neutral nature of inter-species interactions respectively. The inter-species interactions are generally not reciprocal, i.e., $\alpha_{ij} \neq \alpha_{ji}$. We considered nine different models, each denoting a specific type of interspecies interaction (see Table I for the list), e.g., $M_{+-}$ describes the model where the inter-species interactions are given by $\alpha_{12} > 0$ and $\alpha_{21} < 0$. The immune responses are described by monotonically increasing functions with increasing values of $N_1$ and $N_2$ representing concentrations of antimicrobial proteins or neutrophils attracted to the infection site, i.e., $g_1(N_1,N_2)=k_{d1}N_1I$, and, $g_2(N_1,N_2)=k_{d2}N_2I$, where the immune response, I, is generated due to the immune response induced by $N_1$ and $N_2$ , and, is assumed to be additive, i.e., $I=I_1+I_2$, where, $I_1=k_1N_1/(K_{M1}+N_1)$ and $I_2=k_2N_2/(K_{M2}+N_2)$. We write $g_1(N_1,N_2)$ and $g_2(N_1,N_2)$ as,

$g_1(N_1,N_2)=k_{d11}(N_1)^2/(K_{M1}+N_1) + k_{d12}N_1N_2/(K_{M2}+N_2)$, and,

$g_2(N_1,N_2)=k_{d21}N_1N_2/(K_{M1}+N_1) + k_{d22}(N_2)^2/(K_{M2}+N_2)$, where, $k_{dij}=k_{di}k_j$ $(i,j \in \{1,2\})$. Depending on the values of the parameters, the kinetics described by the ODEs in Eq. (1) produce multiple fixed, e.g., $N_1$ is present but $N_2$ is absent, $N_1$ is absent but $N_2$ is present, or both $N_1$ and $N_2$ are present. With appropriate choices of parameter values, these fixed points can become stable fixed points which the system would reach at long times if the initial values are chosen within appropriate ranges (the domain of attraction) (details in the supplementary material). The values of $N_1$ and $N_2$ at the stable fixed points as well as the kinetics of $N_1$ and $N_2$ leading to those fixed points vary as the parameters in the ODEs are changed. Since the parameters in the ODEs (representing the nature of resource utilization, inter- and intra-species interaction, host-immune responses and their effect on the bacterial population) describe the role of the environment and inter-species interactions on size of the bacteria population, we designate these parameters ($\{e_i\}$ and Table II and Table S1) as 'ecological interactions'. We hypothesize that these ecological interactions vary from animal to animal, resulting in different populations of bacterial species infecting/colonizing middle ears in individual animals. Here we address the following questions: (1)What can we deduce about the nature of variations in ecological interactions between individual animals from the experimentally observed variations in bacterial populations? (2) Does the extent of variation of the other ecological interactions depend on the inter-species interactions between the bacterial species? (3) Is it possible that seemingly unrelated ecological interactions are interdependent and this occurs in response to selective pressures on specific bacterial strains in a host population?

*B. MaxEnt formalism to quantify host-host variations:* MaxEnt is widely used in statistical physics[20-22], information theory[23], and statistics[24] to infer distributions of variables based on available measurements; and recently we used MaxEnt to quantify functional implications of cell-to-cell variations of chemotactic protein abundances[25,26].

Here we use MaxEnt to infer the distribution of the ecological interactions in individual animals, using as constraints the observed populations of NTHI and Sp in OM. We introduce a parameter vector, $\{e_i\}$, that represents the parameters in the ODE models and use our MaxEnt based method to estimate the distribution, $\hat{P}(\{e_i\})$. We outline our method for a simple example below and provide further details regarding the full calculation in the supplementary material. The measured values of NTHI (or $N_1$) and Sp (or $N_2$) populations at different time points in single infection (where the middle ear is infected with a single bacterial species) or co-infection (where the middle ear is infected with both NTHI and Sp) experiments provide us with average values and variances of NTHI and Sp populations over an animal population. For example, the average values of the NTHI and Sp populations can be described as,

$$(1/\#\text{ of animals}) \sum_{a=1}^{m} N_{1,2}^{(a)}(t) = \bar{N}_{1,2}^{\text{expt}}(t) = \sum_{\{e_i\}} P(\{e_i\}) N_{1,2}(\{e_i\},t) \qquad (2)$$

where, $N_1^{(a)}(t)$ and $N_2^{(a)}(t)$ refer to the populations $N_1$ and $N_2$ in the middle ear of an animal indexed by *a* at time t (e.g., 7 days after inoculation). Thus, the first equality on the LHS defines the average value of $N_1$ measured at a time t over multiple animals. The second equality on the RHS equates the model values to the experimental measurements. If the ecological niches $\{e_i\}$ are distributed according to a distribution $P(\{e_i\})$ in the animals, and the infection kinetics of $N_1$ and $N_2$ follow the ODEs in Eq. (1), then the average of $N_1(\{e_i\},t)$ and $N_2(\{e_i\},t)$ over $P(\{e_i\})$ should reproduce the observed average value at time t. There are many ways to choose a $P(\{e_i\})$ that will satisfy Eq. (2), we use a MaxEnt based approach that enables us to infer $P(\{e_i\})$ solely based on available data without any additional assumptions. This method selects a $P(\{e_i\})$ that maximizes the Shannon Entropy, $S = -\sum_{\{e_i\}} P(\{e_i\}) \ln P(\{e_i\})$, in the presence of constraints imposed by the available data, such as the second equality in Eq. (2). Instead of directly maximizing S, we estimate $P(\{e_i\})$ by minimizing a relative entropy (Eq. 3). Further details regarding the method are provided in the Methods section and the supplementary material.

**Interspecies interactions regulate animal-to-animal variations of microbial kinetics**

We quantified the extent of variation of the ecological interactions between animals by calculating the minimum value of the relative entropy, MinRE defined in Eq. (3) in Materials and Methods, for all nine models (Fig. 2A). All the models were constrained to reproduce the average values and variances of NTHI and Sp populations measured at 7 days post inoculation when the animals were infected with NTHI and Sp simultaneously. The model in which Sp helps NTHI to access nutrients but NTHI competes with Sp (model $M_{-+}$) produces the smallest MinRE, i.e., this model is consistent with the broadest parameter variations. The next best (MinRE) model was $M_{0+}$, in which Sp stays neutral to NTHI growth and NTHI competes with Sp for resources. In contrast, the models $M_{+-}$ and $M_{0-}$ , in which NTHI helps Sp to access resources, are consistent with only a very small amount of variation in the parameters. The results can be understood in the following way. The experiments show that at 7 days after co-inoculation (NTHI~1000 CFU, Sp~150 CFU), the average population of NTHI (~$10^7$ CFU) is substantially higher than that of Sp (~10 CFU). In contrast, when the animals are infected with either NTHI or Sp alone, both species reach high population (~$10^7$ CFU). Therefore, the models that will produce high growth for NTHI and low growth for Sp at later times (~ 7 days) across a wider range of parameter variations will turn out to be the models with smaller MinRE. In the model $M_{-+}$ , the interspecies interaction supports a higher NTHI and a lower Sp growth since Sp co-operates with NTHI in its growth, but the presence of NTHI counteracts Sp growth. The immune response, regardless of the type of interspecies interaction, can also support a larger NTHI population than Sp population by killing Sp at a higher rate compared to NTHI. However, in the model $M_{-+}$ when the elicited immune response kills NTHI at a higher rate compared to Sp, which can lead to kinetics opposite to those observed in experiments, higher values of the interspecies interactions (e.g., $\alpha_{12}$ and $\alpha_{12}$) can counteract effects induced by the immune response and produce a pattern similar to that observed in experiments. So that the MaxEnt probability distribution is heavily concentrated on the subset of vectors of the ecological interaction parameters for which the immune response is able to counteract this effect and produce higher growth in NTHI compared to Sp. These patterns also indicate how seemingly unrelated ecological interactions, such as the interspecies interactions and the immune response, can become correlated. This is discussed in greater detail in the next section.

Next we compare the nature of variations of the ecological interactions explaining the infection data against the culture experiments (Fig. 2B). In the in vitro culture experiments, both NTHI and Sp grew in the medium when they were inoculated separately or simultaneously (Fig. S2). The population size of NTHI is similar to that of Sp when the bacterial species are cultured individually, and the NTHI population is slightly larger than that of Sp in the co-culture experiments. The models where Sp competes (model $M_{+0}$) or stays neutral (model $M_{00}$) with NTHI for utilizing resources produce the smallest MinREs, whereas the model with only competitive interspecies interactions (model $M_{++}$) produces the highest MinRE. When both species are competing with each other for the common resources, the species can co-exist only within a small range of parameter values, as a small difference between $\alpha_{12}$ and $\alpha_{21}$ can lead to elimination of one species over the other (see the analysis of the ODEs in the supplementary material).  In contrast, when a species is not interacting with another species it always reaches a population size determined by the carrying capacity. Therefore, the models that contain neutral interactions between the species allow for more variation in underlying parameters compared to the other models.

*Testing Predictions*: We used the estimated MaxEnt distributions to generate predictions for measurements that were not used as constraints in fitting the MaxEnt models. Specifically, we predicted the average values of populations of NTHI and Sp at day 3 when the animals were co-inoculated with these species. In addition, we also predicted the correlation between NTHI and Sp at day 7. The predictions from model ($M_{-+}$), which was the best (MinRE) model under the original set of constraints, were in reasonable agreement with the data (Table S2). The models with larger MinRE values produced less agreement with the additional measurements compared to model $M_{+-}$ (Table S2). In general, predictions were better for NTHI than Sp. The disagreement between the model predictions and the data for Sp could point to the importance of spatial structures such as biofilms in regulating the bacterial kinetics. This point is further deliberated in the discussion section.

**Specific ecological interactions become inter-dependent**

In order to further characterize the structure of the MinRE model, we first checked whether the inferred distribution $\hat{P}(\{e_i\})$ of the model parameters could be well approximated by a multivariate normal distribution (Eq. 4), which would imply that average values and pair-correlations between the parameters capture most of the variations in the system. Since the distribution appeared to be well approximated by a multivariate normal distribution (Fig. S3), we quantified the inter-dependencies between the model parameters by using the inverse matrix, $[\Omega]_{ij} = [C^{-1}]_{ij}$, where, $C_{ij}$ denotes the correlation between the model parameters $e_i$ and $e_j$, i.e., $C_{ij} = \overline{(e_i - \mu_i)(e_j - \mu_j)}$ and $\mu_i = \overline{e_i}$ ; where the over bar indicates the average over $\hat{P}(\{e_i\})$. We further quantified the strength of the interdependence or relationship between the model parameters by calculating a metric ($\{[\mathrm{Int}]_{ij}\}$) for any pair of parameters $e_i$ and $e_j$ using the Ω matrix (see the Methods section for details). A larger magnitude of $[\mathrm{Int}]_{ij}$ implies a greater contribution of the pair of parameters to determining animal-to-animal (or trial-to-trial) variations (Fig. 3), and a negative or a positive value indicates whether the members of the pair vary in the same or opposite direction while keeping the response unchanged.

Analysis of the inter-dependencies using $\{[\mathrm{Int}]_{ij}\}$ for $M_{-+}$ showed (Fig. 3A) that parameters not directly related to the immune response, such as the carrying capacity or the strength of interspecies interactions, became dependent on parameters directly related to the immune response, such as the rate of killing of Sp by the immune system. The number of such dependencies with higher magnitudes of $\{[\mathrm{Int}]_{ij}\}$increases for the higher MinRE models (consistent with less variation in the parameters) (Fig. 3B, D) in the niches considered here. This result can be intuitively understood as follows: the requirement of having more interdependence between the parameters imposes greater restrictions on the sets of parameter vectors that are able to reproduce the measured average values. The majority of the dependencies can be explained qualitatively or by analyzing the ODEs. E.g., the increase in the NTHI bacterial load required to induce the maximum immune response that favors an increase in the NTHI population is compensated for by the corresponding decrease in the available resources or the carrying capacity (Fig. 3A). This implies that in order to be consistent with the experimental data, a particular strain in NTHI that is less efficient in stimulating the immune response will also undergo changes that reduce its capability to utilize the nutrients. Further explanations regarding the other interdependencies are provided in the supplementary material (Table S3).

These in vitro analyses show (Fig. 3C,D) that parameters describing inter-species interactions, in contrast to intra-species interactions, become more dependent during infections, e.g., a decrease in the carrying capacity for Sp, which would support a higher population of NTHI due to lower competitive interspecies interaction, is compensated by a decrease in the strength in interspecies competition between NTHI and Sp.

**Discussion**

We developed a MaxEnt based method to quantify host-to-host variations of ecological interactions for two bacterial species, NTHI and Sp, which are responsible for polymicrobial OM infection. A key finding of this analysis is the dependency of the extent of host-to-host variations of the ecological interactions on the nature of the underlying bacterial inter-species interactions. Cooperative-competitive or neutral-competitive interaction models between bacterial species allow for the largest variations (smaller values of MinRE) of the ecological interactions or model parameters, and are likely to be associated with host populations with greater heterogeneity and environmental perturbations. Interspecies interactions between NTHI and Sp arise via a range of processes, such as secretion of toxins, metabolic byproducts, inflammation and quorum sensing. Since the nature and magnitude of these processes can vary from strain to strain in a bacterial species, it is possible that specific strains of NTHI and Sp possessing interspecies interactions, ones that can accommodate the largest variations in ecological niches in the host population, are selected as the host immune system and the microorganisms co-evolve in an evolutionary arms race[5].

The structure of the inferred variations of ecological interactions reveal that seemingly unrelated ecological variables, such as the carrying capacities and the host immune response, become interdependent. For example, in the model $M_{-+}$, which shows the largest variation in the ecological interactions, a mutually co-operative relationship between carrying capacity and the rate of bacteria elimination would imply selection of an NTHI strain that can use the same resources more efficiently, whose growth is also better suppressed by the host immune response (Fig. 3). These results suggest that if the microbial communities residing in the host have the flexibility to accommodate changes in the ecological interactions, for example, by altering gene expressions[27], these changes are likely to occur in a coordinated manner[28].

The in vitro culture experiments show that both the NTHI and Sp bacterial strains are able to co-exist in the culture medium, but in the chinchilla host, the Sp strains are eliminated in the presence of NTHI. This clearly suggests a qualitative difference in ecological niches for growth in the host microenvironment and the in vitro culture. Our MaxEnt based analysis quantitatively characterizes the difference. Our analysis showed that the neutral model ($M_{+0}$) produced a wider spread in ecological interactions in vitro over the purely competitive model ($M_{++}$). This result is consistent with Gause's law in population dynamics[7], which states that two species competing for the same resources cannot co-exist. In contrast, in the presence of the host immune response, the purely competitive model ($M_{++}$) showed a much wider variation compared to the neutral model ($M_{00}$). Furthermore, the models associated with the largest MinRE values *in vitro* and in the host are composed of very different interspecies interactions . These differences emphasize the importance of the immune response in manipulating ecological niches and evolutionary selection of bacterial strains residing in the host.

We primarily studied models that approximated and simplified interspecies interactions in terms of a relatively small number of parameters. Therefore, these models need to be modified in order to investigate interspecies interactions such as quorum sensing, which increases fitness of the same strain, or the formation of spatial structures such as biofilms, which help bacterial species to evade the host immune response. The importance of these effects, in particular biofilm formation, becomes apparent as the predictions from the two-species models differ from the measurements of abundances of bacterial populations at higher inoculation doses. The two-species model could be extended to include additional strains associated with biofilms found in the chinchilla middle ear[2]. Investigation of the role of these additional strains in host-to-host variations of infection kinetics would be an interesting future direction.

Our analysis showed that host-to-host variations of polymicrobial infection kinetics can provide valuable clues regarding evolutionary selection of bacterial strains and the role of the host immune response in shaping the fitness landscape of the polymicrobial community. A possible test of the results presented here could be analysis of gene expression from bacterial isolates obtained from the middle ear pre- and post-co-inoculation. If genes responsible for metabolization of essential metals are expressed further during the course of infection in NTHI but not in Sp, this would lend further support to our conclusion that specific attributes helping in NTHI growth are selected due to the combined effect of the presence of Sp and the host immune response. However, the modeling approach proposed here represents a general method, not limited to OM, which can be utilized to understand mechanisms of host-microorganism relationships and their evolutionary origin using measurements delineating host-to-host variations of microbial and host response kinetics.

**Methods and Materials:**

Solution of the ODEs: The ODEs in Eq. (2) were solved using the software package BIONETGEN[29]. The codes used in the simulations can be found at http://planetx.nationwidechildrens.org/~jayajit/.

Estimation of $\hat{P}(\{e_i\})$: We used measurements from infection and culture experiments studying kinetics of single or two bacterial species for estimating $P(\{e_i\})$. We separate the parameter vector $\{e_i\}$ into two sub-sets $\{e_i^{(S)}\}$ and $\{e_i^{(M)}\}$ (see Table II and Table S1) that represent respectively the parameters solely regulating bacterial kinetics for experiments with single species and the additional parameters required to describe the kinetics for the mixed co-infection/culture experiments. We described the kinetics in terms of dimensionless parameters $\{\tilde{e}_i\}$ constructed (Table S1 and supplementary material) from constructed $\{e_i\}$ and carried out all the MaxEnt analysis on the dimensionless parameters. Thus, $\{e_i\}$ in the rest of the section refer to $\{\tilde{e}_i\}$. We retained the same symbols for simplicity. $\hat{P}(\{e_i\})$ can be decomposed into $\hat{P}(\{e_i\}) = \hat{P}^{(M)}(\{e_i\})\hat{P}^{(S)}(\{e_i^S\},\{0\})$ (see the supplementary material for the derivation), where, $\hat{P}^{(M)}(\{e_i\})$ and $\hat{P}^{(S)}(\{e_i^S\},\{0\})$ describe the distributions of the parameters consistent with experiments done with single or two bacterial species, respectively. We briefly describe the numerical scheme used in estimating $\hat{P}^{(S)}(\{e_i^S\},\{0\})$ and $\hat{P}^{(M)}(\{e_i\})$.

*(A) Infection experiments*: $\hat{P}^{(S)}(\{e_i^S\},\{0\})$ is estimated from the infection experiments where the chinchilla middle ears are infected with either Sp or NTHI. The a priori distribution of the parameters before the maximization of S was assumed to be a uniform distribution in $\{e_i^{(S)}\}$ as the uniform distribution represents the maximally uncertain state of a system and the parameters related to mixed two species experiments set to zero, i.e., $q_U(\{e_i\}) = q_U(\{e_i^{(S)}\}) \times \prod_i \delta_{e_i^{(M)},0}$, where, $\delta_{ab}$=1 (or =0)when a=b (or a≠b). We constrained the average values of populations of NTHI and Sp according to their measured values at two different times (3 and 7 days). $P(\{e_i^{(S)}\},\{0\})$ is estimated by minimizing the relative entropy

$$MinRE^{(S)} = \sum_{\{e_i^{(S)}\}} P(\{e_i^{(S)}\},\{0\}) \ln[P(\{e_i^{(S)}\},\{0\}) / q_U(\{e_i^{(S)}\},\{0\})]$$

subject to the contraints imposed by the average values.

In the next step, we generate the a priori distribution $q(\{e_i\})$ by choosing parameters $\{e_i^{(S)}\}$ based on $\hat{P}^{(S)}(\{e_i^S\},\{0\})$ and the parameters $\{e_i^{(M)}\}$ were chosen from a uniform distribution, i.e., $q^{(M)}(\{e_i\}) = \hat{P}^{(S)}(\{e_i^{(S)}\},\{0\}) \times q_U(\{e_i^{(M)}\})$. Then we estimate the distribution, $P^{(M)}(\{e_i\})$ when $\{e_i^{(M)}\}$ are not vanishing using the measured values from the co-infection experiments as constraints and minimizing the relative entropy,

$$MinRE = \sum_{\{e_i\}} P^{(M)}(\{e_i\}) \ln[P^{(M)}(\{e_i\}) / q_U(\{e_i\})] \quad (3)$$

,where, $q_U(\{e_i\})$ denotes a uniform distribution for parameters in both the subsets $\{e_i^{(S)}\}$ and $\{e_i^{(M)}\}$. The details regarding sample size and the sampling method are given in the supplementary material.

*(B) In vitro culture*: Since the immune response is absent in the culture experiments, we set $g_1=g_2=0$ in the models. The growth of NTHI and Sp are described by the rates, $f_1(N_1,N_2) = r_1[N_1^2/(K_{lag1}+N_1^2)](K_1-\alpha_{11}N_1-\alpha_{12}N_2)$ and $f_2(N_1,N_2) = r_2[N_2^2/(K_{lag2}+N_2^2)](K_2-\alpha_{22}N_2-\alpha_{21}N_1)$. The terms $[N_1^2/(K_{lag1}+N_1^2)]$ and $[N_2^2/(K_{lag2}+N_2^2)]$ describe the initial lag in the growth of NTHI and Sp. The rest of the parameters are described in the same manner as in the infection models. The distribution of the parameters is estimated using the same scheme as described above from the in vitro measurements studying growth of NTHI and Sp growing individually or simultaneously in the medium.

*Experimental techniques*: Streptococcus pneumoniae TIGR4 and H. influenzae 86-028NP were cultured, alone or together in equivalent ratios, in brain-heart infusion (Difco) supplemented with hemin and NAD, and containing 10% horse serum (HemoStat Laboratories), essentially as described previously[30]. Bacterial counts were derived by plate-count.

*Quantification of the relationship between the model parameters*: We approximate the distribution $\hat{P}^{(M)}(\{e_i\})$ by a multivariate normal distribution (Fig. S3), i.e.,

$$\hat{P}^{(M)}(\{e_i\}) \propto e^{-\sum_{i,j}(e_i-\mu_i)\Omega_{ij}(e_j-\mu_j)} \quad (4)$$

,where, $\{\mu_i\}$ denote the average values of the parameters $\{e_i\}$, and, $[\Omega^{-1}]_{ij} = C_{ij}$ ; $C_{ij}$ denoting the correlation between the niches $e_i$ and $e_j$ or $C_{ij} = \overline{(e_i-\mu_i)(e_j-\mu_j)}$ where the overbar indicates the average over $\hat{P}^{(M)}(\{e_i\})$. The elements of the matrix $\Omega$ demonstrate the "interaction" between the parameters or the nature of the relationship between the parameters in producing the observed correlations. E.g., a positive (or negative) value $\Omega_{ij}$ would imply the parameters $e_i$ and $e_j$ counter-act (or help) each other in producing the observed population kinetics. A vanishing value of $\Omega_{ij}$ would imply very little relationship between $e_i$ and $e_j$. We evaluated which of the interactions in ($\{\Omega_{ij}\}$) contribute the most in determining the observed covariance $C_{ij}$. This was done by not constraining a specific $C_{ij}$, and then comparing the inferred $\hat{P}^{*(ij)}(\{e_i\})$ with the original inferred distribution, $\hat{P}(\{e_i\})$ using the Kullback-Leibler distance,

$[D_{KL}]_{ij} = \sum_{\{e_i\}} \hat{P}(\{e_i\}) \ln[\hat{P}(\{e_i\}) / \hat{P}^{*(ij)}(\{e_i\})]$. A larger $[D_{KL}]_{ij}$ implies a greater contribution of a particular $\Omega_{ij}$ in determining the animal-to-animal variations of the ecological niches (Fig. 3). Therefore, we use a metric, $Int_{ij}=sgn(\Omega_{ij})\ [D_{KL}]_{ij}$, to quantify inferred interaction strength between the pair of niches, i and j.

**Acknowledgements:**

We are grateful to Lauren Bakaletz for a critical reading of the manuscript. The work is supported by a grant from the NIH to WES, WCR, VJV, CJ, and, JD. JD is also partially supported by the Research Institute at the Nationwide Children's Hospital and a grant from the Ohio Supercomputer Center (OSC).

**List of Tables**

**Table I: List of the models considered.**

| Effect on growth | $M_{++}$ | $M_{+-}$ | $M_{+0}$ | $M_{-+}$ | $M_{--}$ | $M_{-0}$ | $M_{0+}$ | $M_{0-}$ | $M_{00}$ |
|---|---|---|---|---|---|---|---|---|---|
| NTHI on Sp | + | + | + | - | - | - | 0 | 0 | 0 |
| Sp on NTHI | + | - | 0 | + | - | 0 | + | - | 0 |

+ = counteracts, - = helps, 0 = stays neutral

**Table II: Parameters involved in single and two bacterial species experiments.**

| | $\{e^{(S)}_i\}$ | $\{e^{(M)}_i\}$ |
|---|---|---|
| Infection | $K_1$, $K_2$, $\alpha_{11}$, $\alpha_{22}$, $k_{d11}$, $K_{M1}$, | $\alpha_{12}$, $\alpha_{21}$, $k_{d12}$, $k_{d21}$ |
| | $k_{d22}$, $K_{M2}$ | |
| *In vitro* culture | $K_1$, $K_2$, $K_{lag1}$, $K_{lag2}$, $\alpha_{11}$, $\alpha_{22}$, | $\alpha_{12}$, $\alpha_{21}$ |

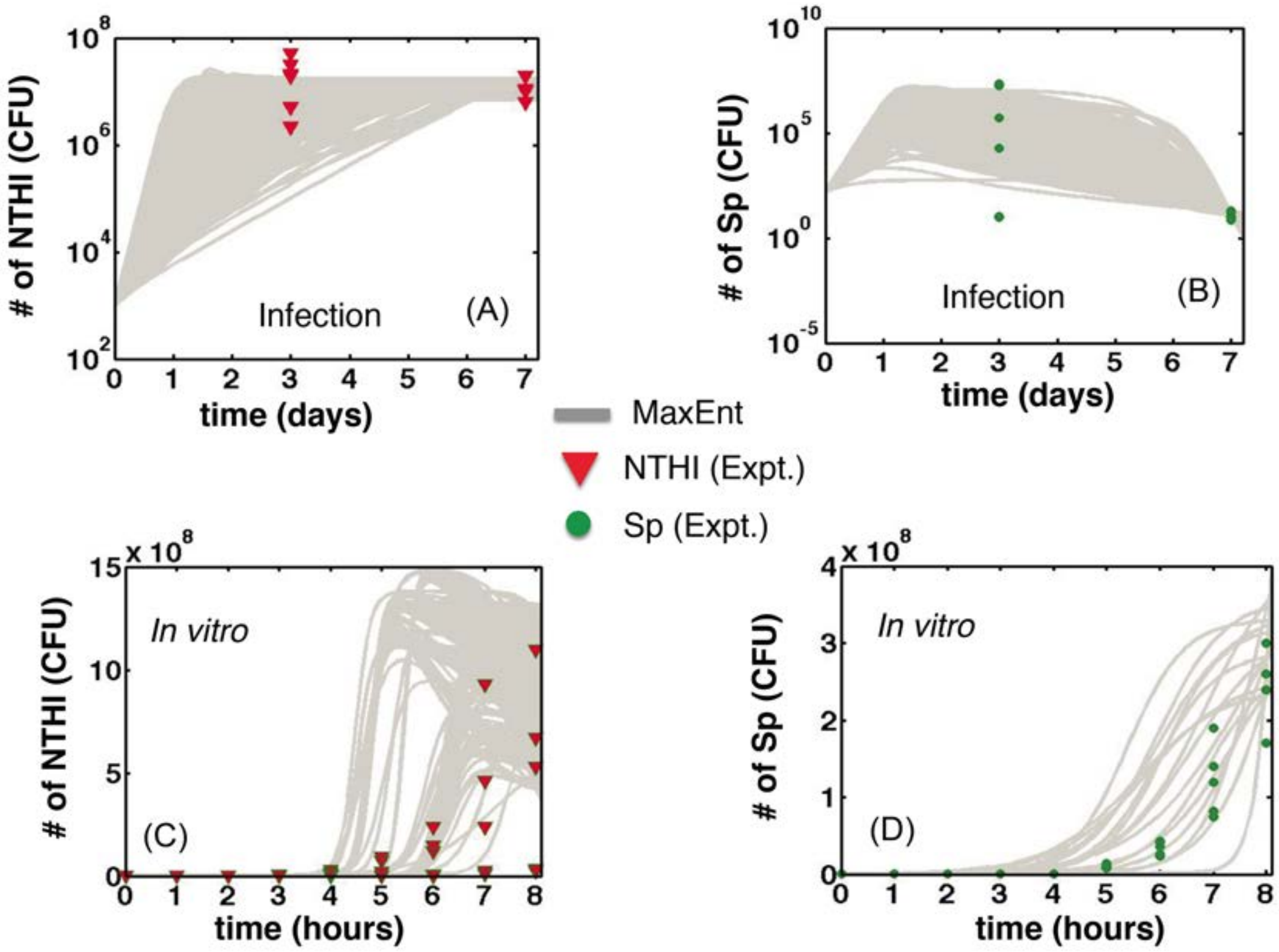

**Fig. 1 Variations of bacterial kinetics between hosts and culture medium trials.** (A) Kinetics of NTHI population (grey lines) the middle ears of individual in "silico animals" when the animals were co-inoculated with NTHI and Sp. The animals were distributed according to the inferred MaxEnt distribution for the $M_{+-}$ model. The experimental data (each red triangle corresponds to an individual chinchilla middle ear) were taken from Ref.[2] where the animals received inocula of $\sim 10^3$ CFU and ~150 CFU of NTHI and Sp, respectively. (B) Kinetics of Sp for the same set up as in (A) and shown using the same visualization scheme. (C) Kinetics of NTHI (grey lines) in individual trials drawn from the inferred MaxEnt distribution for the $M_{+0}$ model in silico co-culture of NTHI and Sp. The experimental data for each trials for the co-culture experiments with NTHI and Sp

are shown in red triangles. (D) Kinetics of Sp for the same set up as in (C).

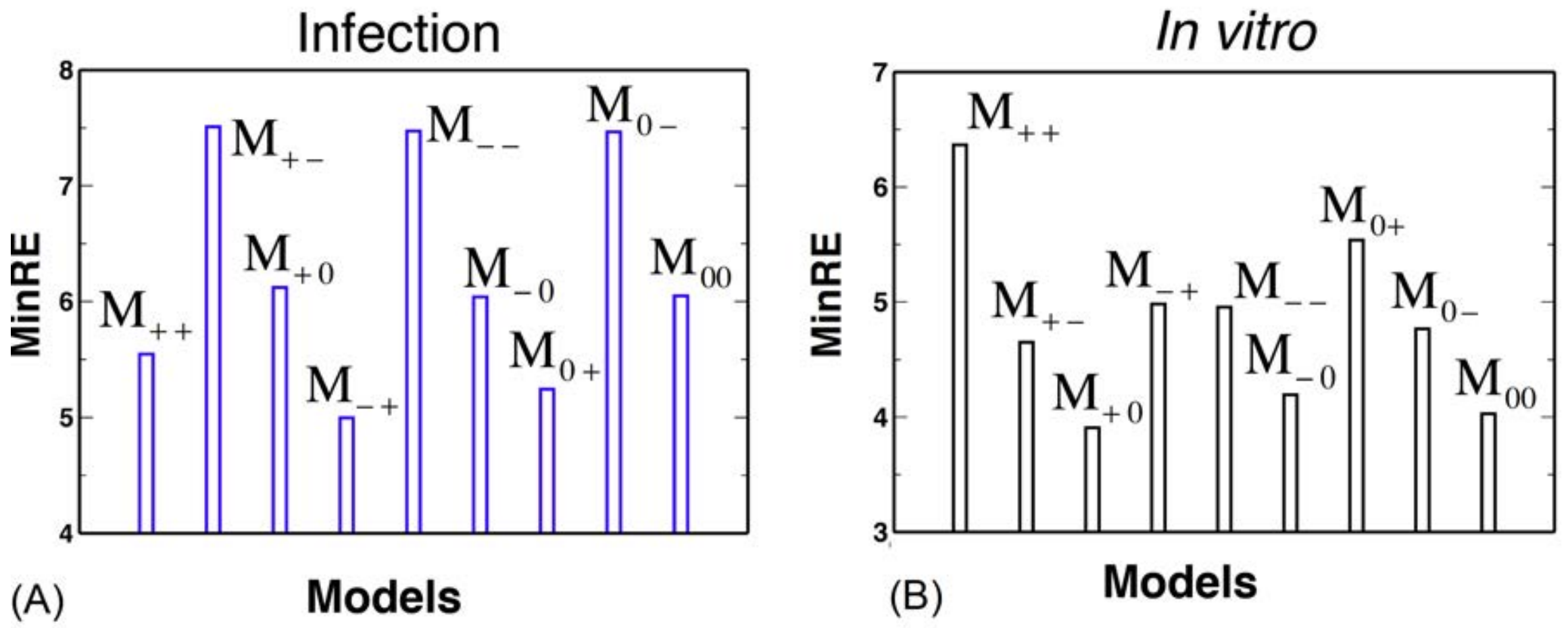

**Fig. 2 Interspecies ecological interactions regulate variations of bacterial kinetics.** MinRE values, quantifying the extent of variations of ecological interactions, show differences in the abilities to accommodate individual-to-individual variances in models containing qualitatively different types of interspecies ecological interactions in an animal population (A) or a set of trials in culture experiments (B).

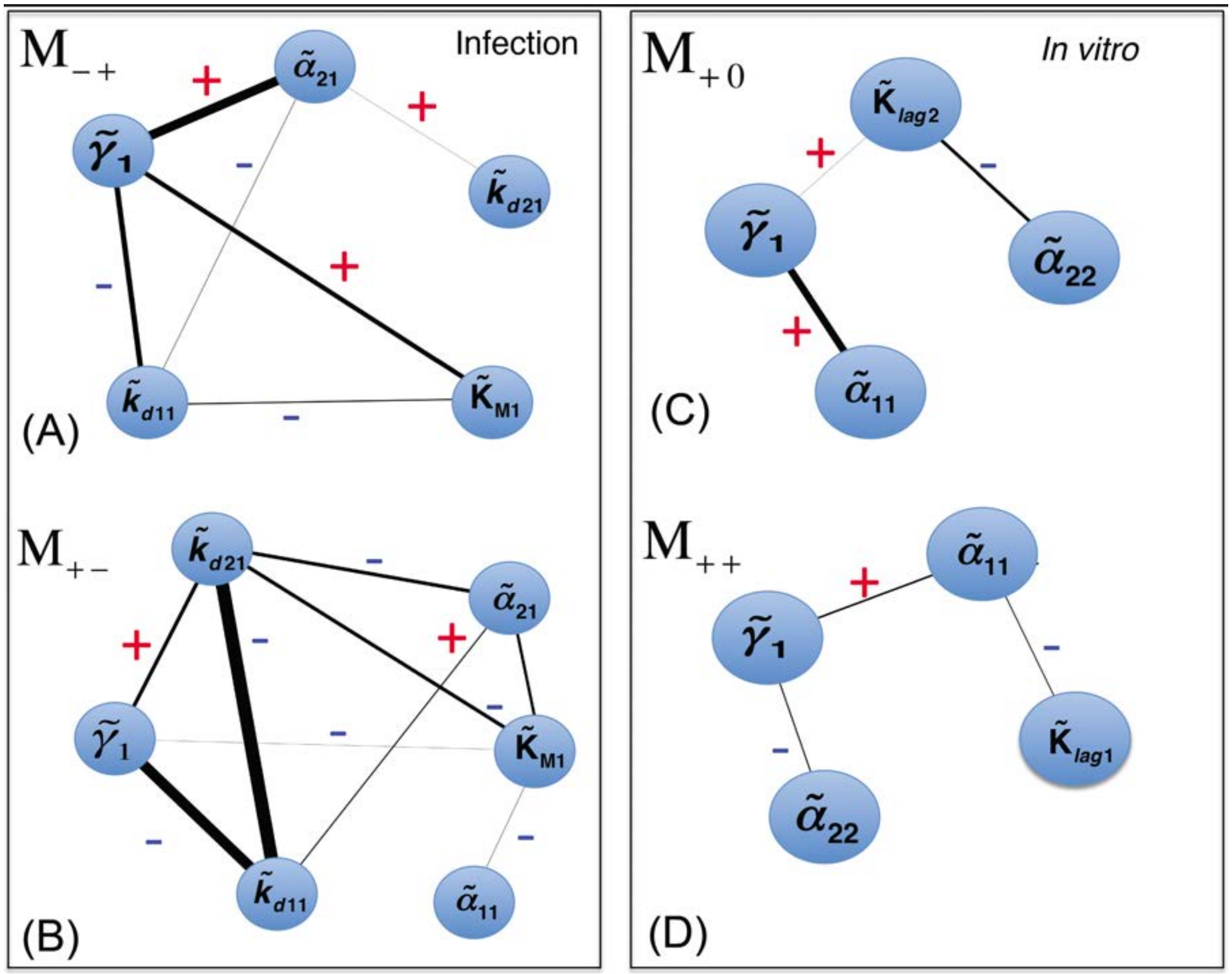

**Fig. 3 Characteristics of the inferred distribution of the ecological interactions.** (A) Inter-dependencies between ecological interactions (shown in terms of the dimensionless parameters shown in Table S1) described by the metric $[Int]_{ij}$ for the model $M_{-+}$ (smallest MinRE) for the infection data. Higher $[Int]_{ij}$ values are shown with thicker lines and $[Int]_{ij}$ values less than the threshold (abs$[[Int]_{ij}]> 0.1$) are now shown. The +ve and the –ve signs are also indicted. (B) Same as in (A) for the model $M_{+-}$ for the infection data which produce the largest MinRE value. (C) $[Int]_{ij}$ shown for the model $M_{+0}$ (smallest MinRE) explaining the culture data. (D) Same as in (C) for the model $M_{++}$ (largest MinRE).

**Supplementary Material for "Host-to-host variation of ecological interactions in polymicrobial infections"**

## Section 1: Fixed points and the stability of the ODE models

### 1.1: In vitro model and the conditions for stability

The ODEs for the growth of NTHI ($N_1$) and Sp ($N_2$) populations in in vitro culture are given by,

$$\frac{dN_1}{dt} = \frac{N_1^2}{\mathrm{K}_{lag1} + N_1^2}\left(\gamma_1 N_1 - \alpha_{11} N_1^2 - \alpha_{12} N_1 N_2\right)$$
$$\frac{dN_2}{dt} = \frac{N_2^2}{\mathrm{K}_{lag2} + N_2^2}\left(\gamma_2 N_2 - \alpha_{22} N_2^2 - \alpha_{21} N_1 N_2\right) \qquad \text{S1}$$

where γ' s and the α's represent the doubling rate of the bacteria and the inter and intra species competition for resource respectively. Using the same scheme for notation as in the main text, the parameters are defined as, $\gamma_1 = r_1 K_1$, $\gamma_2 = r_2 K_2$, $\alpha_{11} = r_1\alpha_{11}$(main text), $\alpha_{12} = r_1\alpha_{12}$(main text), $\alpha_{21} = r_2\alpha_{21}$(main text) and $\alpha_{22} = r_2\alpha_{22}$(main text) respectively. $K_{lag1}$ and $K_{lag2}$ are the parameters that determine the respective lags in the bacterial growth observed in in vitro experiments. For $K_{lag1}$, $K_{lag2} >> N_1^0$, $N_2^0$ the initial growth of the bacteria is stunted. Only after $N_1$ and $N_2$ overcome the thresholds set by $K_{lag1}$ and $K_{lag2}$, the bacteria can transition to an exponentially growing phase. The system described by Eqn (S1) allows for four fixed points, namely, 1. $\{N_1^s = 0,\ N_2^s = 0\}$, 2. $\{N_1^s = \gamma_1/\alpha_{11},\ N_2^s = 0\}$, 3. $\{N_1^s = 0,\ N_2^s = \gamma_2/\alpha_{22}\}$ and 4. $\{N_1^s = \alpha_{22}\gamma_1 - \alpha_{12}\gamma_2/|\alpha|,\ N_2^s = \alpha_{11}\gamma_2 - \alpha_{21}\gamma_1/|\alpha|\}$, where $|\alpha| = \alpha_{11}\alpha_{22} - \alpha_{12}\alpha_{21}$.

The stability of the fixed points can be studied by analyzing the stability matrix $S$ given by

$$S = \begin{pmatrix} \frac{(N_1^s)^2}{\mathrm{K}_{lag1} + (N_1^s)^2}\left[\gamma_1 - 2\alpha_{11}N_1^s - \alpha_{12}N_2^s\right] & \frac{-\alpha_{12}(N_1^s)^3}{\mathrm{K}_{lag1} + (N_1^s)^2} \\ \frac{-\alpha_{21}(N_2^s)^3}{\mathrm{K}_{lag2} + (N_2^s)^2} & \frac{(N_2^s)^2}{\mathrm{K}_{lag2} + (N_2^s)^2}\left[\gamma_2 - 2\alpha_{22}N_2^s - \alpha_{21}N_1^s\right] \end{pmatrix}$$

Assuming $\gamma_1,\ \gamma_2 > 0$, we can see that the first fixed point ($N_1^s$=$N_2^s$=0) is unstable if we take the third derivative of the RHS of Eqn S1. The second fixed point is stable in the $N_1$ direction while the third fixed point is only stable in the $N_2$ direction as can be seen by substituting the steady state values of $N_1$ and $N_2$ in the stability matrix $S$. When both NTHI and Sp co-exist the characteristic equation (λ being the eigenvalue) can be written as

$$\left|\begin{pmatrix} a\left[\gamma_1 - 2\alpha_{11}N_1^s - \alpha_{12}N_2^s\right] - \lambda & -a\alpha_{12}N_1^S \\ -b\alpha_{21}N_2^S & b\left[\gamma_2 - 2\alpha_{22}N_2^s - \alpha_{21}N_1^s\right] - \lambda \end{pmatrix}\right| = 0$$

where $a = \frac{\left(N_1^s\right)^2}{\mathrm{K}_{lag1} + \left(N_1^s\right)^2} > 0$ and $b = \frac{\left(N_2^s\right)^2}{\mathrm{K}_{lag2} + \left(N_2^s\right)^2} > 0$. For $N_1^{\mathrm{s}}$ and $N_2^{\mathrm{s}}$ greater than $\mathrm{K}_{\mathrm{lag1}}$ and $\mathrm{K}_{\mathrm{lag2}}$, respectively, (See Fig S2) we have a=b~1.

Let us for the time being let us assume that $\gamma_1 \sim \gamma_2$ (the doubling rate for NTHI is roughly one hour whereas the doubling rate of Sp is half an hour). Then the eigenvalues are given by

$$\lambda_{1,2} = -\gamma_1,\ \frac{\gamma_1\left(\alpha_{11} - \alpha_{21}\right)\left(\alpha_{12} - \alpha_{22}\right)}{\alpha_{11}\alpha_{22} - \alpha_{12}\alpha_{21}}$$

The conditions for the stability of the co-existence fixed point are summarized below.

**No interaction (model $M_{00}$)**

1. In the absence of interspecies interaction, we can see that the fixed point is always stable. The two species evolve irrespective of one another.

**Pure competition (model $M_{++}$)** $\left(\alpha_{12} > 0,\ \alpha_{21} > 0\right)$

1. If $0 < \alpha_{12} < \alpha_{22}, 0 < \alpha_{21} < \alpha_{11}$, the fixed point is stable.
2. If $0 < \alpha_{12} > \alpha_{22}, 0 < \alpha_{21} < \alpha_{11}$, the fixed point is unstable.
3. If $0 < \alpha_{12} < \alpha_{22}, 0 < \alpha_{21} > \alpha_{11}$, the fixed point gets unstable.
4. If $0 < \alpha_{12} > \alpha_{22}, 0 < \alpha_{21} > \alpha_{11}$, the fixed point is unstable.

Therefore, for a competition model, the co-existence phase is only stable provided the interspecies coupling is small compared to the intra species coupling.

**Co-operation/competition (model $M_{-+}$)** $\left(\alpha_{12} < 0,\ \alpha_{21} > 0\right)$

1. If $\alpha_{12} < 0, 0 < \alpha_{21} < \alpha_{11}$, the fixed point is stable.
2. If $\alpha_{12} < 0, 0 < \alpha_{21} > \alpha_{11}$, the fixed point is unstable.

For a co-operation/competition model, the co-existence is guaranteed provided the interspecies competition is small compared to the intra species competition. The analysis for $\alpha_{21} < 0$ is similar.

**Pure co-operation (model $M_{--}$)** $\left(\alpha_{12} < 0,\ \alpha_{21} < 0\right)$

1. If $\alpha_{12} < 0, \alpha_{21} < 0$, the fixed point is stable provided $\alpha_{11}\alpha_{22} > \alpha_{12}\alpha_{21}$.

**Neutral competition (model $M_{0+}$)** ($\alpha_{12} = 0,\ \alpha_{21} > 0$)

1. If $0 < \alpha_{21} < \alpha_{11}$, the fixed point is stable.
2. If $0 < \alpha_{21} > \alpha_{11}$, the fixed point is unstable.

Like the competition model, as long as the inter species coupling is small compared to the intra species competition we have co-existence. The model $M_{+0}$ can be analyzed in a similar fashion.

**Neutral co-operation (model $M_{0-}$)** ($\alpha_{12} = 0,\ \alpha_{21} < 0$)

1. The co-existence fixed point is always stable. The model $M_{-0}$ is the same way.

In a nutshell, as long as the inter species competition for resource is not fierce we can have a stable co-existence.

## 1.2: Infection model and the conditions for stability

The kinetics of two species co-infection is modeled using the ODEs given by

$$\frac{dN_1}{dt} = \left(\gamma_1 N_1 - \alpha_{11} N_1^2 - \alpha_{12} N_1 N_2\right) - \frac{k_{d11} N_1^2}{K_{M1} + N_1} - \frac{k_{d12} N_1 N_2}{K_{M2} + N_2}$$

$$\frac{dN_2}{dt} = \left(\gamma_2 N_2 - \alpha_{22} N_2^2 - \alpha_{21} N_1 N_2\right) - \frac{k_{d21} N_1 N_2}{K_{M1} + N_1} - \frac{k_{d22} N_2^2}{K_{M2} + N_2} \qquad \text{S2}$$

where the $\gamma$' s and the $\alpha$'s are defined in the same way as the previous section. We will focus our attention to the steady state relevant for us (see Weimer et al. Fig 1), namely $N_1^s \neq 0,\ N_2^s = 0$

The steady state for $N_1$ is

$$\gamma_1 - \alpha_{11} N_1^s - \frac{k_{d11} N_1^s}{K_{M1} + N_1^s} = 0$$

$$\Rightarrow N_1^s = \frac{-\left(k_{d11} + \alpha_{11} K_{M1} - \gamma_1\right) + \sqrt{\left(k_{d11} + \alpha_{11} K_{M1} - \gamma_1\right)^2 + 4\gamma_1 \alpha_{11} K_{M1}}}{2\alpha_{11}} \qquad \text{S3}$$

The stability matrix $S$ is given by

$$S = \begin{pmatrix} \gamma_1 - 2\alpha_{11} N_1^s - \dfrac{k_{d11} N_1^S (2K_{M1} + N_1^S)}{\left(K_{M1} + N_1^S\right)^2} & -\alpha_{12} N_1^S - \dfrac{k_{d12} N_1^S}{K_{M2}} \\ 0 & \gamma_2 - \alpha_{21} N_1^s - \dfrac{k_{d21} N_1^S}{K_{M1} + N_1^S} \end{pmatrix}$$

The eigenvalues of $S$ are given by,

$$\lambda_{1,2} = \gamma_1 - 2\alpha_{11}N_1^s - \frac{k_{d11}N_1^S(2K_{M1}+N_1^S)}{\left(K_{M1}+N_1^S\right)^2},\ \gamma_2 - \alpha_{21}N_1^s - \frac{k_{d21}N_1^S}{K_{M1}+N_1^S} \qquad \text{S4}$$

We have looked at two possible scenarios.

**Case I: The decay rates** $k_{d11}, k_{d21} \approx \varepsilon < \gamma_1$. The immune response plays a marginal role in this case. $\alpha_{21}$ drives the extinction of $N_2$. The steady state in Eqn (S3) can be recast as

$$N_1^S = \frac{-(\alpha_{11}K_{M1}-\gamma_1)-\varepsilon+(\alpha_{11}K_{M1}+\gamma_1)\sqrt{1+\frac{2\varepsilon(\alpha_{11}K_{M1}-\gamma_1)}{(\alpha_{11}K_{M1}+\gamma_1)^2}}}{2\alpha_{11}}$$

$$= \frac{-(\alpha_{11}K_{M1}-\gamma_1)-\varepsilon+(\alpha_{11}K_{M1}+\gamma_1)\left(1+\frac{\varepsilon(\alpha_{11}K_{M1}-\gamma_1)}{(\alpha_{11}K_{M1}+\gamma_1)^2}\right)}{2\alpha_{11}}$$

$$= \frac{2\gamma_1-\varepsilon\left(1-\frac{(\alpha_{11}K_{M1}-\gamma_1)}{(\alpha_{11}K_{M1}+\gamma_1)}\right)}{2\alpha_{11}} = \frac{\gamma_1}{\alpha_{11}}\left(1-\frac{\varepsilon}{\alpha_{11}K_{M1}+\gamma_1}\right) \qquad \text{S5}$$

Substituting Eqn (S5) in the expression of $\lambda_1$ in Eqn (S4) we have

$$\gamma_1 - 2\gamma_1\left(1-\frac{\varepsilon}{\alpha_{11}K_{M1}+\gamma_1}\right) - \frac{\varepsilon\frac{\gamma_1}{\alpha_{11}}\left(1-\frac{\varepsilon}{\alpha_{11}K_{M1}+\gamma_1}\right)\left(2K_{M1}+\frac{\gamma_1}{\alpha_{11}}\left(1-\frac{\varepsilon}{\alpha_{11}K_{M1}+\gamma_1}\right)\right)}{\left(K_{M1}+\frac{\gamma_1}{\alpha_{11}}\left(1-\frac{\varepsilon}{\alpha_{11}K_{M1}+\gamma_1}\right)\right)^2}$$

The condition of stability demands that both $\lambda_1$ and $\lambda_2 < 0$. Thus to $O(\varepsilon)$ $\lambda_1 < 0$ implies

$$-\gamma_1 + \frac{2\gamma_1\varepsilon}{\alpha_{11}K_{M1}+\gamma_1} - \frac{\varepsilon\gamma_1(2K_{M1}\alpha_{11}+\gamma_1)}{(K_{M1}\alpha_{11}+\gamma_1)^2} \approx -\gamma_1 + \frac{2\gamma_1\varepsilon}{\alpha_{11}K_{M1}+\gamma_1} - \frac{\varepsilon\gamma_1}{\alpha_{11}K_{M1}+\gamma_1} = -\gamma_1 + \frac{\gamma_1\varepsilon}{\alpha_{11}K_{M1}+\gamma_1} < 0$$

$$\Rightarrow \varepsilon = k_{d11} < \alpha_{11}K_{M1} + \gamma_1 \qquad \text{S6}$$

For the expression of $\lambda_2$ to $O(\varepsilon)$ we have

$$\gamma_2 < \alpha_{21} N_1^S + \frac{\varepsilon N_1^S}{K_{M1} + N_1^S}$$

$$\gamma_2 < \alpha_{21}\frac{\gamma_1}{\alpha_{11}} - \frac{\varepsilon \gamma_1 \alpha_{21}}{\alpha_{11}(\alpha_{11} K_{M1} + \gamma_1)} + \frac{\varepsilon \gamma_1}{\alpha_{11} K_{M1} + \gamma_1}$$

$$\gamma_2 < \alpha_{21}\frac{\gamma_1}{\alpha_{11}} + \frac{\varepsilon \gamma_1 (\alpha_{11} - \alpha_{21})}{\alpha_{11}(\alpha_{11} K_{M1} + \gamma_1)} \qquad \text{S7}$$

The extinction of $N_2$ is mainly driven by the interspecies competition $\alpha_{21}$. As long as $\gamma_2 < \alpha_{21}\frac{\gamma_1}{\alpha_{11}}$ and Eqn (S6) are simultaneously satisfied we have a stable fixed point.

**Case II:** $\gamma_1, k_{d21} >> k_{d11}, \alpha_{21} \approx \varepsilon$. In the absence of a strong interspecies interaction the immune response drives the decay of $N_2$.

The condition for stability (Eqn (S4)) dictates that

$$\gamma_2 < \frac{\gamma_1 \varepsilon}{\alpha_{11}}\left(1 - \frac{\varepsilon}{K_{M1}\alpha_{11} + \gamma_1}\right) + \frac{k_{d21}\frac{\gamma_1}{\alpha_{11}}\left(1 - \frac{\varepsilon}{K_{M1}\alpha_{11} + \gamma_1}\right)}{K_{M1} + \frac{\gamma_1}{\alpha_{11}}\left(1 - \frac{\varepsilon}{K_{M1}\alpha_{11} + \gamma_1}\right)}$$

To O(ε) we have

$$\gamma_2 < \frac{k_{d21}\gamma_1}{K_{M1}\alpha_{11} + \gamma_1}\left(1 - \frac{K_{M1}\alpha_{11}\varepsilon}{(K_{M1}\alpha_{11} + \gamma_1)^2}\right) + \frac{\gamma_1 \varepsilon}{\alpha_{11}} \qquad \text{S8}$$

## 1.3: The condition for the existence of a maximum in Sp kinetics as observed by Weimer et al.

The results in Weimer et al. (1) showed that the average population of NTHI grows and saturates while the average population of Sp increases initially and then falls back to an undetectable range by 7 days. In the following section we delineate the condition for a transient growth of Sp followed by its decay. We can rewrite Eqn (S2) as

$$\frac{dN_2}{dt} = N_2 \Xi(N_1, N_2) \text{ where } \Xi(N_1, N_2) = \gamma_2 - \alpha_{21}N_1 - \alpha_{22}N_2 - \frac{k_{d21}N_1}{K_{M1} + N_1} - \frac{k_{d22}N_2}{K_{M2} + N_2}$$

The fact that $N_2$ reaches a maximum value ($N_2^{\max}$) at a finite time ($t=t_m$) and then starts decaying demands that $dN_2/dt\,|_{t=tm} = 0$ and $d^2N_2/dt^2|_{t=tm} < 0$ and $N_2=N_2^{\max}$, which yields

$\Xi(N_1, N_2) = 0$ and $N_2 \partial_{N_1}\Xi(N_1, N_2)\frac{dN_1}{dt} < 0$ at $N_2 = N_2^{\max}$. Now

$$\partial_{N_1}\Xi(N_1, N_2) = -\alpha_{21} - \frac{k_{d21}K_{M1}}{\left(K_{M1} + N_1(t_m)\right)^2}.$$

The second term in the expression above is always negative. So as long as $N_1$ competes with $N_2$ for resource ($\alpha_{21} > 0$) the derivative above is negative. Thus $dN_1/dt$ has to be greater than zero at $t_m$ for $N_2$ to have a maximum. For a competition model, the growth in Sp is arrested and reversed owing to the growth in NTHI, which elicits an immune response killing Sp at a rate of $k_{d21}N_1/(K_{M1}+N_1)$. In order to get the transient kinetics in $N_2$ in a co-operative model ($\alpha_{21} < 0$), either $k_{d21}$ or $N_1(t_m)$ or both have to be much larger such that,

$$-\alpha_{21} < \frac{k_{d21}K_{M1}}{\left(K_{M1}+N_1(t_m)\right)^2}.$$

## Section 2: Derivation of the MaxEnt distribution

We provide details regarding the MaxEnt scheme that we use to estimate the distribution $\hat{P}(\{e_i\})$. The parameters, $\{e_i\}$, determine the kinetics of $N_1$ and $N_2$ in the ODEs (Eq. 1) used to explain the observed NTHI and Sp mono infection and co-infection. The parameters, $\{e_i\}$, can be decomposed into two subsets (see Table S1), $\{e_i^{(S)}\}$ and $\{e_i^{(M)}\}$, that represent the parameters required to describe experiments with single bacteria species and the additional parameters required to describe the kinetics in the mixed co-infection/culture experiments, respectively. The constraints on the distribution $P(\{e_i\})$ are imposed by the 3 days and 7 days average population of NTHI and Sp for single species inoculation and by the 7 days average values and variances of populations of NTHI and Sp for co-inoculation. We show the derivation for a smaller set of constraints. The calculations can be easily generalized.

Let us assume that we know the average values of NTHI and Sp measured at 7 days post infection in single bacteria and two bacteria experiments. Therefore, the constraints for the single bacteria experiments are given by,

$$\begin{aligned}
&\sum_{\{e_i^S\}} P(\{e_i^S\},\{0\}) N_1^{(S)}(\{e_i^S\},\{0\},t=7d) = \bar{N}_1^{(S)\text{expt}}(t=7d) \\
&\sum_{\{e_i^S\}} P(\{e_i^S\},\{0\}) N_2^{(S)}(\{e_i^S\},\{0\},t=7d) = \bar{N}_2^{(S)\text{expt}}(t=7d)
\end{aligned} \qquad \text{(S9)}$$

where, $N_{1,2}^{(S)}(\{e_i^{(S)}\},\{0\},t=7d)$ refer to the abundances of NTHI and Sp at $t = 7$ days calculated from the ODEs when the parameters $\{e_i^{(M)}\}$ are set to zero. $P(\{e^{(S)}\})$ denotes the distribution of $\{e_i^{(S)}\}$ when $\{e_i^{(M)}\}$ are set to zero. $\bar{N}_{1,2}^{(S)\text{expt}}(t=7d)$ indicate the average values of NTHI and Sp calculated at 7 days in experiments. Similarly, the constraints for the two bacteria species experiments are given by,

$$\begin{aligned}
&\sum_{\{e_i^{(S)}\},\{e_i^{(M)}\}\notin\{0\}} P(\{e_i^{(S)}\},\{e_i^{(M)}\}) N_1^{(M)}(\{e_i^{(S)}\},\{e_i^{(M)}\},t=7d) = \bar{N}_1^{(M)\text{expt}}(t=7d) \\
&\sum_{\{e_i^{(S)}\},\{e_i^{(M)}\}\notin\{0\}} P(\{e_i^{(S)}\},\{e_i^{(M)}\}) N_2^{(M)}(\{e_i^{(S)}\},\{e_i^{(M)}\},t=7d) = \bar{N}_2^{(M)\text{expt}}(t=7d)
\end{aligned}$$

(S10)

where, $\{e_i^{(M)}\}$ are *not* equal to zero. The subscript (M) denotes the values in the mixed two species ODE solutions or experiments.

We maximize the entropy, $S = -\sum_{\{e_i\}} P(\{e_i\}) \ln P(\{e_i\})$, subject to the above constraints and the normalization constraint, $\sum_{\{e_i\}} P(\{e_i\}) = 1$ . Therefore, the estimated distribution $\hat{P}(\{e_i\})$ can be obtained from the equation given below.

$$\delta S - \lambda_1^{(S)} \sum_{\{e_i^{(S)}\}} \delta P(\{e_i^{(S)}\},\{0\}) N_1^{(S)} - \lambda_2^{(S)} \sum_{\{e_i^{(S)}\}} \delta P(\{e_i^{(S)}\},\{0\}) N_2^{(S)}$$

$$-\lambda_1^{(M)} \sum_{\{e_i^{(S)}\},\{e_i^{(M)}\}\notin\{0\}} \delta P(\{e_i^{(S)}\},\{e_i^{(M)}\}) N_1^{(M)} - \lambda_2^{(M)} \sum_{\{e_i^{(S)}\},\{e_i^{(M)}\}\notin\{0\}} \delta P(\{e_i^{(S)}\},\{e_i^{(M)}\}) N_2^{(M)} - \lambda_3 \sum_{\{e_i\}} \delta P(\{e_i\}) = 0$$

$$\Rightarrow -\sum_{\{e_i\}} \delta P(\{e_i\})(\ln P(\{e_i\})+1) - \left[ \lambda_1^{(S)} \sum_{\{e_i\}} \delta P(\{e_i^{(S)}\},\{0\}) N_1^{(S)} + \lambda_2^{(S)} \sum_{\{e_i\}} \delta P(\{e_i^{(S)}\},\{0\}) N_2^{(S)} \right] (\prod_j \delta_{e_j^{(M)},0})$$

$$-\left[ \lambda_1^{(M)} \sum_{\{e_i\}} \delta P(\{e_i^{(S)}\},\{e_i^{(M)}\}) N_1^{(M)} + \lambda_2^{(M)} \sum_{\{e_i\}} \delta P(\{e_i^{(S)}\},\{e_i^{(M)}\}) N_2^{(M)} \right] \left[ 1 - \prod_j \delta_{e_j^{(M)},0} \right] - \lambda_3 \sum_{\{e_i\}} \delta P(\{e_i\}) = 0$$

(S11)

The solution of the above equation is,

$$\hat{P}(\{e_i\}) \propto \exp\left[ -\left(\lambda_1^{(S)} N_1^{(S)} + \lambda_2^{(S)} N_2^{(S)}\right)\left(\prod_j \delta_{e_j^{(M)},0}\right) - \left(\lambda_1^{(M)} N_1^{(M)} + \lambda_2^{(M)} N_2^{(M)}\right)\left(1 - \prod_j \delta_{e_j^{(M)},0}\right) - \lambda_3 \right]$$

(S12)

The dependence of $\{e_i^{(S)}\}$ and $\{e_i^{(M)}\}$ on $\hat{P}(\{e_i\})$ arises through the variation of $N_{1,2}^{(M)}(\{e_i^{(S)}\},\{e_i^{(M)}\},t=7d)$ with respect to nonzero $\{e_i^{(M)}\}$ and $\{e_i^{(S)}\}$. The terms proportional to $N_{1,2}^{(S)}(\{e_i^{(S)}\},\{0\},t=7d)$ generate variations of $\hat{P}(\{e_i\})$ on $\{e_i^{(S)}\}$ only when all the parameters in $\{e_i^{(M)}\}$ are set to zero. Therefore, we can decompose, $\hat{P}(\{e_i\}) = \hat{P}^{(M)}(\{e_i\})\hat{P}^{(S)}(\{e_i^S\},\{0\})$, where,

$$\hat{P}^{(M)}(\{e_i\}) \propto \exp\left[ -\left(\lambda_1^{(M)} N_1^{(M)} + \lambda_2^{(M)} N_2^{(M)}\right)\left(1 - \prod_j \delta_{e_j^{(M)},0}\right) \right] \quad \text{(S13)}$$

and

$$\hat{P}^{(S)}(\{e_i^{(S)}\},\{0\}) \propto \exp\left[ -\left(\lambda_1^{(S)} N_1^{(S)} + \lambda_2^{(S)} N_2^{(S)}\right)\left(\prod_j \delta_{e_j^{(M)},0}\right) \right] \quad \text{(S14)}$$

Since the interspecies interactions are described by $\{e_i^{(M)}\} \notin \{0\}$, all the models considered will have the same dependence on $\{e_i^{(S)}\}$ via $\hat{P}^{(S)}(\{e_i^S\},\{0\})$ . Therefore, the differences in the variations of the ecological niches in the models will be given by $\hat{P}^{(M)}(\{e_i\})$ and we quantify variations for each model by the relative entropy,

$$MinRE = \sum_{\{e_i^{(S)}\},\{e_i^{(M)}\}\notin\{0\}} \hat{P}^{(M)}(\{e_i\})\ln[P^{(M)}(\{e_i\})/q_U(\{e_i\})] \quad . \qquad \text{(S15)}$$

where $q_U(\{e_i\})$ is the uniform distribution over all the parameters.

## Section 3: Numerical scheme to evaluate the MaxEnt distributions

### A: Infection models

The rate equations Eqn (S2) for the mixed infection can be recast in a dimensionless form given below

$$\frac{d\tilde{N}_1}{d\tau} = \tilde{\gamma}_1\tilde{N}_1 - \tilde{N}_1^2 - \tilde{\alpha}_{12}\tilde{N}_1\tilde{N}_2 - \frac{\tilde{k}_{d11}\tilde{N}_1^2}{\tilde{K}_{M1}+\tilde{N}_1} - \frac{\tilde{k}_{d12}\tilde{N}_1\tilde{N}_2}{\tilde{K}_{M2}+\tilde{N}_2}$$
$$\frac{d\tilde{N}_2}{d\tau} = \tilde{\gamma}_2\tilde{N}_2 - \tilde{N}_2^2 - \tilde{\alpha}_{21}\tilde{N}_1\tilde{N}_2 - \frac{\tilde{k}_{d21}\tilde{N}_1\tilde{N}_2}{\tilde{K}_{M1}+\tilde{N}_1} - \frac{\tilde{k}_{d22}\tilde{N}_2^2}{\tilde{K}_{M2}+\tilde{N}_2}$$

(S16)

where $\tau=(\gamma_1+\gamma_2)t$, $\tilde{\gamma}_1=\gamma_1/(\gamma_1+\gamma_2)$, $\tilde{\gamma}_2=1-\tilde{\gamma}_1$, $\tilde{\alpha}_{11}=\alpha_{11}/(\gamma_1+\gamma_2)$, $\tilde{\alpha}_{22}=\alpha_{22}/(\gamma_1+\gamma_2)$, $\tilde{N}_1=\tilde{\alpha}_{11}N_1$, $\tilde{N}_2=\tilde{\alpha}_{22}N_2$, $\tilde{\alpha}_{12}=\alpha_{12}/\alpha_{22}$, $\tilde{\alpha}_{21}=\alpha_{21}/\alpha_{11}$.
$\tilde{k}_{d11}$, $\tilde{k}_{d12}$, $\tilde{k}_{d21}$, $\tilde{k}_{d22}$ are obtained by dividing the respective $k_d$'s in Eqn(S2) by $\gamma_1+\gamma_2$.
$\tilde{K}_{M1}=K_{M1}\tilde{\alpha}_{11}$ and $\tilde{K}_{M2}=K_{M2}\tilde{\alpha}_{22}$.

*Evaluation of the MaxEnt distribution for the single species kinetics data*

For single species inoculation (either by $10^3$ CFU of NTHI or by 150 CFU of Sp) the inter species interaction parameters $\tilde{\alpha}_{12}$, $\tilde{\alpha}_{21}$, $\tilde{k}_{d12}$, $\tilde{k}_{d21}$ are set to zero. $\gamma_1$, $\gamma_2$, $\tilde{K}_{M1}$, $\tilde{K}_{M2}$, $\tilde{k}_{d11}$, $\tilde{k}_{d22}$ are chosen from a uniform distribution U(0,10), where, U(a,b) denotes a normalized uniform distribution between a and b. Upon drawing $\gamma_1$, $\gamma_2$ the corresponding values of $\tilde{\gamma}_1$, $\tilde{\gamma}_2$ are calculated. Values of $N_1^0$ and $N_2^0$ are set to 1000 and 150 CFU respectively. The intraspecies competition parameters $\tilde{\alpha}_{11}$ and $\tilde{\alpha}_{22}$ are varied uniformly in a window of [$1.2\text{x}10^{-10}$, $1.6\text{x}10^{-8}$]. Upon drawing all the eight parameters we solve the ODEs in Eqn (S9) and read out the values of $N_1$ and $N_2$ at 3 and 7 days respectively. Each tuple of eight parameters, referred to as $\{e_i^{(S)}\}$, represents an animal in our simulation. We have drawn these eight parameter tuples for 100,000 times in order to simulate a cohort of 100,000 animals.

We sought a $P^{(S)}(\{e^{(S)}\},\{0\})$ (for convenience we will drop the $\{0\}$ in the argument and denote the probability distribution as $P^{(S)}(\{e^{(S)}\})$) that will maximize the Shannon Entropy

$$S = -\sum_{\{e^{(S)}\}} P\left(\left\{e^{(S)}\right\}\right)\ln\left(P\left(\left\{e^{(S)}\right\}\right)\right) \text{ with the constraints}$$

$$\sum_{\{e^{(s)}\}} N_1(\{e^{(s)}\},t=3d)P(\{e^{(s)}\}) = \bar{N}_1^{\text{expt}}(t=3d) = 1.61\times10^7$$
$$\sum_{\{e^{(s)}\}} N_2(\{e^{(s)}\},t=3d)P(\{e^{(s)}\}) = \bar{N}_2^{\text{expt}}(t=3d) = 3.4\times10^6$$
$$\sum_{\{e^{(s)}\}} N_1(\{e^{(s)}\},t=7d)P(\{e^{(s)}\}) = \bar{N}_1^{\text{expt}}(t=7d) = 4.66\times10^7$$
$$\sum_{\{e^{(s)}\}} N_2(\{e^{(s)}\},t=7d)P(\{e^{(s)}\}) = \bar{N}_2^{\text{expt}}(t=7d) = 3.4\times10^6$$

(S17)

(Note the time units are in days and the RHS is in CFU)

Maximizing Shannon entropy with the constraints in Eqn (S17) yields

$$\hat{P}\left(\left\{e^{(S)}\right\}\right) = Z^{-1}\exp\left(-\lambda_1^{(S)}N_1^{(S)}\left(\left\{e^{(S)}\right\},3\right) - \lambda_2^{(S)}N_2^{(S)}\left(\left\{e^{(S)}\right\},3\right) - \lambda_3^{(S)}N_1^{(S)}\left(\left\{e^{(S)}\right\},7\right) - \lambda_4^{(S)}N_2^{(S)}\left(\left\{e^{(S)}\right\},7\right)\right)$$

(S18)

where $Z$ is the partition sum and $\{\lambda^{(S)}\}$ are the Lagrange multipliers. We plug in the expression for $\hat{P}\left(\left\{e^{(S)}\right\}\right)$ in Eqn (S17) and solve for the $\{\lambda^{(S)}\}$.

*Evaluation of the MaxEnt distribution for the co-infection kinetics data*

We draw the animals (tuple of eight parameters) that are most likely to yield the single species inoculation data from the MaxEnt distribution $\hat{P}\left(\left\{e^{(S)}\right\}\right)$. Upon drawing we inoculate that animal with both the bacteria simultaneously by assigning non zero values to the interaction parameters $\tilde{\alpha}_{12}$, $\tilde{\alpha}_{21}$, $\tilde{k}_{d12}$, $\tilde{k}_{d21}$. The values are drawn from a uniform distribution U(0,10). We then solve ODEs in Eqn (S9). The values of mixed species $N_1$ and $N_2$ referred to as $N_1^M$ and $N_2^M$ respectively, are read out at 3 and 7 days. We constrain the mean and the second moments of NTHI and Sp abundances at 7 days.

The constraints used are given below.

$$\sum_{\{e_i^{(S)}\},\{e_i^{(M)}\}\notin\{0\}} P(\{e_i^{(S)}\},\{e_i^{(M)}\}) N_1^{(M)}(\{e_i^{(S)}\},\{e_i^{(M)}\}, t=7d) = \bar{N}_1^{(M)\text{expt}}(t=7d) = 1.2\times 10^7$$

$$\sum_{\{e_i^{(S)}\},\{e_i^{(M)}\}\notin\{0\}} P(\{e_i^{(S)}\},\{e_i^{(M)}\}) N_1^{2\ (M)}(\{e_i^{(S)}\},\{e_i^{(M)}\}, t=7d) = \bar{N}_1^{2(M)\text{expt}}(t=7d) = 1.5\times 10^{14}$$

$$\sum_{\{e_i^{(S)}\},\{e_i^{(M)}\}\notin\{0\}} P(\{e_i^{(S)}\},\{e_i^{(M)}\}) N_2^{(M)}(\{e_i^{(S)}\},\{e_i^{(M)}\}, t=7d) = \bar{N}_2^{(M)\text{expt}}(t=7d) = 13$$

$$\sum_{\{e_i^{(S)}\},\{e_i^{(M)}\}\notin\{0\}} P(\{e_i^{(S)}\},\{e_i^{(M)}\}) N_2^{2\ (M)}(\{e_i^{(S)}\},\{e_i^{(M)}\}, t=7d) = \bar{N}_2^{2(M)\text{expt}}(t=7d) = 1.84\times 10^2$$

(S19)

where $P(\{e_i^{(S)}\},\{e_i^{(M)}\})$ is the MaxEnt distribution over all the parameters and the RHS is in CFU.

## B. In vitro culture models

Eqn (S1) can be recast in the dimensionless form given below

$$\frac{d\tilde{N}_1}{d\tau} = \frac{\tilde{N}_1^2}{\tilde{K}_{lag1} + \tilde{N}_1^2}\left(\tilde{\gamma}_1 \tilde{N}_1 - \tilde{N}_1^2 - \tilde{\alpha}_{12}\tilde{N}_1\tilde{N}_2\right)$$

$$\frac{d\tilde{N}_2}{d\tau} = \frac{\tilde{N}_2^2}{\tilde{K}_{lag2} + \tilde{N}_2^2}\left(\tilde{\gamma}_2 \tilde{N}_2 - \tilde{N}_2^2 - \tilde{\alpha}_{21}\tilde{N}_1\tilde{N}_2\right)$$

(S20)

where $\tilde{K}_{lag1}$ and $\tilde{K}_{lag2}$ are $\tilde{\alpha}_{11}^2 K_{lag1}$ and $\tilde{\alpha}_{22}^2 K_{lag2}$ respectively. The rest of the tilde variables are same as the one defined before in Eqn (S16).

*Evaluation of the MaxEnt distribution for the single species kinetics data*

For single species culture the inter species interaction parameters $\tilde{\alpha}_{12}$, $\tilde{\alpha}_{21}$ are set to zero. The initial populations are set to 1.0x10$^6$ for NTHI and 4.5x10$^5$ for Sp. These numbers are obtained by averaging the three experimental trials shown in Fig (S2). Like the infection model, $\gamma_1$, $\gamma_2$ are chosen from a uniform distribution U(0,10) and then the dimensionless variables, $\tilde{\gamma}_1$ and $\tilde{\gamma}_2$ are calculated. $\tilde{K}_{lag1}$, $\tilde{K}_{lag2}$ are drawn from U(3.6x10$^7$,3.6x10$^5$) and U(1.0x10$^8$, 5.0x10$^6$) respectively, while $\alpha_{11}$ and $\alpha_{22}$ are drawn from U(4.7x10$^{11}$, 4.7x10$^9$) and U(3.1x10$^{-11}$, 3.1x10$^9$) respectively. Upon drawing all the six parameters we solve the ODEs in Eqn (S20) and read out the values of $N_1$ and $N_2$ at 4 and 8 hours respectively. Each tuple of six parameters, referred to as $\{e^{(S)}\}$, represents an experiment trial in our simulation. We have simulated 100,000 such trials.

As in the case of the infection models, we constructed a maximum entropy distribution $P\left(\left\{e^{(S)}\right\}\right)$, of the form

$$P\left(\left\{e^{(S)}\right\}\right)=Z^{-1}\exp\left(-\lambda_1^{(S)}N_1^{(S)}\left(\left\{e^{(S)}\right\},4\right)-\lambda_2^{(S)}N_2^{(S)}\left(\left\{e^{(S)}\right\},4\right)-\lambda_3^{(S)}N_1^{(S)}\left(\left\{e^{(S)}\right\},8\right)-\lambda_4^{(S)}N_2^{(S)}\left(\left\{e^{(S)}\right\},8\right)\right) \tag{S21}$$

that respects the following constraints.

$$\begin{aligned}
&\sum_{\{e^{(S)}\}} N_1\left(\left\{e^{(S)}\right\},t=4\right)P\left(\left\{e^{(S)}\right\}\right)=\overline{N}_1^{\text{expt}}(t=4)=6.43\times10^6\\
&\sum_{\{e^{(S)}\}} N_2\left(\left\{e^{(S)}\right\},t=4\right)P\left(\left\{e^{(S)}\right\}\right)=\overline{N}_2^{\text{expt}}(t=4)=1.613\times10^6\\
&\sum_{\{e^{(S)}\}} N_1\left(\left\{e^{(S)}\right\},t=8\right)P\left(\left\{e^{(S)}\right\}\right)=\overline{N}_1^{\text{expt}}(t=8)=5.9\times10^8\\
&\sum_{\{e^{(S)}\}} N_2\left(\left\{e^{(S)}\right\},t=8\right)P\left(\left\{e^{(S)}\right\}\right)=\overline{N}_2^{\text{expt}}(t=8)=2.96\times10^8
\end{aligned} \tag{S22}$$

(Note the time units are in hours and the RHS is in CFU)

*Evaluation of the MaxEnt distribution for kinetics in co-culture experiments*

We use the MaxEnt distribution in Eqn (S21) to draw the most likely experiments. Then we introduce co-culture interaction by drawing numbers for $\tilde{\alpha}_{12}$, $\tilde{\alpha}_{21}$ from a uniform distribution U(0,10). Then we rerun the ODEs given by Eqn (S20) and read out the values of $N_1$ and $N_2$ at 4, 6 and 8 hours. Like the infection model, we constrain the mean and variances of the abundance of NTHI and Sp at 8 hours. The constraints are given by,

$$\sum_{\{e_i^{(S)}\},\{e_i^{(M)}\}\notin\{0\}} P(\{e_i^{(S)}\},\{e_i^{(M)}\}) N_1^{(M)}(\{e_i^{(S)}\},\{e_i^{(M)}\},t=8h) = \bar{N}_1^{(M)\text{expt}}(t=8h) = 5.75\times 10^8$$

$$\sum_{\{e_i^{(S)}\},\{e_i^{(M)}\}\notin\{0\}} P(\{e_i^{(S)}\},\{e_i^{(M)}\}) N_1^{2(M)}(\{e_i^{(S)}\},\{e_i^{(M)}\},t=8h) = \bar{N}_1^{2\ (M)\text{expt}}(t=8h) = 5.25\times 10^{17}$$

$$\sum_{\{e_i^{(S)}\},\{e_i^{(M)}\}\notin\{0\}} P(\{e_i^{(S)}\},\{e_i^{(M)}\}) N_2^{(M)}(\{e_i^{(S)}\},\{e_i^{(M)}\},t=8h) = \bar{N}_2^{(M)\text{expt}}(t=8h) = 2.61\times 10^8$$

$$\sum_{\{e_i^{(S)}\},\{e_i^{(M)}\}\notin\{0\}} P(\{e_i^{(S)}\},\{e_i^{(M)}\}) N_2^{2\ (M)}(\{e_i^{(S)}\},\{e_i^{(M)}\},t=8h) = \bar{N}_2^{2\ (M)\text{expt}}(t=8h) = 7.06\times 10^{16}$$

(S23)

The RHS is in CFU. Following the same prescription as in the infection models we calculate the MaxEnt distribution for the co-culture experiments.

**Table S1: List of parameters that are varied in single and two bacterial species experiments.**

| | $\{e^{(S)}_i\}$ | $\{e^{(M)}_i\}$ |
|---|---|---|
| Infection experiments | $\gamma_1$, $\gamma_2$, $\tilde{\alpha}_{11}$, $\tilde{\alpha}_{22}$, $\tilde{k}_{d11}$, $\tilde{K}_{M1}$, $\tilde{k}_{d22}$, $\tilde{K}_{M2}$ | $\tilde{\alpha}_{12}$, $\tilde{\alpha}_{21}$, $\tilde{k}_{d12}$, $\tilde{k}_{d21}$ |
| In vitro culture experiments | $\gamma_1$, $\gamma_2$, $\tilde{K}_{lag1}$, $\tilde{K}_{lag2}$, $\tilde{\alpha}_{11}$, $\tilde{\alpha}_{22}$, | $\tilde{\alpha}_{12}$, $\tilde{\alpha}_{21}$ |

The relation between the dimensionless parameters and the original parameters are shown in Section 1.

**Table S2: Prediction using the inferred distributions for the best and the worst (MinRE) models.**

Predictions for mean values ($\mu_1=\overline{N_1}$, $\mu_2=\overline{N_2}$) variances ($\sigma_1^2=\overline{N_1^2}-\mu_1^2$, $\sigma_2^2=\overline{N_2^2}-\mu_2^2$), covariances ($\text{Cov}(ij)=\overline{N_i N_j}-\overline{N_i}\,\overline{N_j}$, $i\neq j$), and correlations ($\rho_{ij}=\text{Cov}(ij)/\sigma_i\sigma_j$, $i\neq j$) between abundances of NTHI ($N_1$) and Sp ($N_2$).

The best and the worst models are choosen based on their MinRE scores (See Fig 2 main text). For the in vivo experiment the best model in $M_{-+}$ whereas the worst model is $M_{+-}$. For in vitro culture experiment the best model in $M_{+0}$ and the worst model is $M_{++}$.

| | Comparison | $\mu_1$ | $\mu_2$ | $\sigma_1^2$ | $\sigma_2^2$ | Cov(12) | $\rho_{12}$ |
|---|---|---|---|---|---|---|---|
| **Infection** | **Prediction** model $M_{-+}$ (3 days) | $9.04 \times 10^6$ | $5.72 \times 10^5$ | $2.72 \times 10^{13}$ | $1.58 \times 10^{12}$ | $-2.8 \times 10^{12}$ | -0.42 |
| | **Prediction** model $M_{+-}$ (3 days) | $1.06 \times 10^7$ | $3.98 \times 10^4$ | $1.9 \times 10^{13}$ | $2.81 \times 10^{10}$ | $-1.95 \times 10^{11}$ | -0.264 |
| | **Experiment** (3 days) | $2.18 \times 10^7$ | $7.05 \times 10^6$ | $2.82 \times 10^{14}$ | $9.8 \times 10^{13}$ | $9.6 \times 10^{13}$ | 0.57 |
| | **Prediction** model $M_{-+}$ (7 days) | Constraints used in MaxEnt | | | | $-9.3 \times 10^5$ | -0.046 |
| | **Prediction** model $M_{+-}$ (7 days) | | | | | $1.98 \times 10^5$ | 0.01 |
| | **Experiment** (7 days) | | | | | $-3.5 \times 10^5$ | -0.018 |
| ***in vitro* culture** | **Prediction** model $M_{+0}$ (6 hrs) | $7.57 \times 10^8$ | $4.54 \times 10^7$ | $3.95 \times 10^{17}$ | $2.29 \times 10^{15}$ | $-0.8 \times 10^{16}$ | -0.27 |
| | **Prediction** model $M_{++}$ (6 hrs) | $3.28 \times 10^8$ | $1.32 \times 10^8$ | $2.61 \times 10^{17}$ | $7.95 \times 10^{16}$ | $-2.6 \times 10^{16}$ | -0.18 |
| | **Experiment** (6 hrs) | $1.26 \times 10^8$ | $3.28 \times 10^7$ | $9.5 \times 10^{15}$ | $0.64 \times 10^{14}$ | $-0.26 \times 10^{15}$ | -0.31 |

**Table S3: Analysis of the interdependencies in Fig. 3**

| **Model ($M_{-+}$)** | **Parameter1** | **Parameter2** | **Explanation of the observed dependency** |
|---|---|---|---|
| | $\tilde{\gamma}_1$ | $\tilde{k}_{d11}$ | +ve dependency. If $k_{d11}$ increases $\gamma_1$ needs to increase in order to keep $N_1^S$ unchanged (Eqn S5) and for the stability of the steady state (Eqn (S6)) as well. |
| | $\tilde{\gamma}_1$ | $\tilde{K}_{M1}$ | -ve dependency. As $\gamma_1$ increases $K_{M1}$ needs to decrease in order to keep $N_1^S$ unchanged. (According to Eqn S2, lower $K_{M1}$ elicits a stronger and quicker immune response) |
| | $\tilde{\gamma}_1$ | $\tilde{\alpha}_{21}$ | -ve dependency. With an increase |
| | | | in $\gamma_1$ a lower value of $\alpha_{21}$ is required to satisfy Eqn S19 |
| | $\tilde{k}_{d11}$ | $\tilde{\alpha}_{21}$ | + ve dependency. An increase in $k_{d11}$ tends to eliminate $N_1$, the effect of which can be counteracted by an increase in $\alpha_{21}$ which in turn ascertains the decay of $N_2$. (Eqn S5 and Eqn S7) |
| | $\tilde{k}_{d11}$ | $\tilde{K}_{M1}$ | + ve dependency. In order to keep the steady state of $N_1$ fixed, an increase in $k_{d11}$ has to be accompanied by an increase in $K_{M1}$ (Eqn S5) and also to render the steady state stable (Eqn S6). |
| | $\tilde{\alpha}_{21}$ | $\tilde{k}_{d21}$ | -ve dependency. With an increase |
| | | | in $\alpha_{21}$ we only require smaller values of $k_{d21}$ to eliminate $N_2$ at 7 days. |

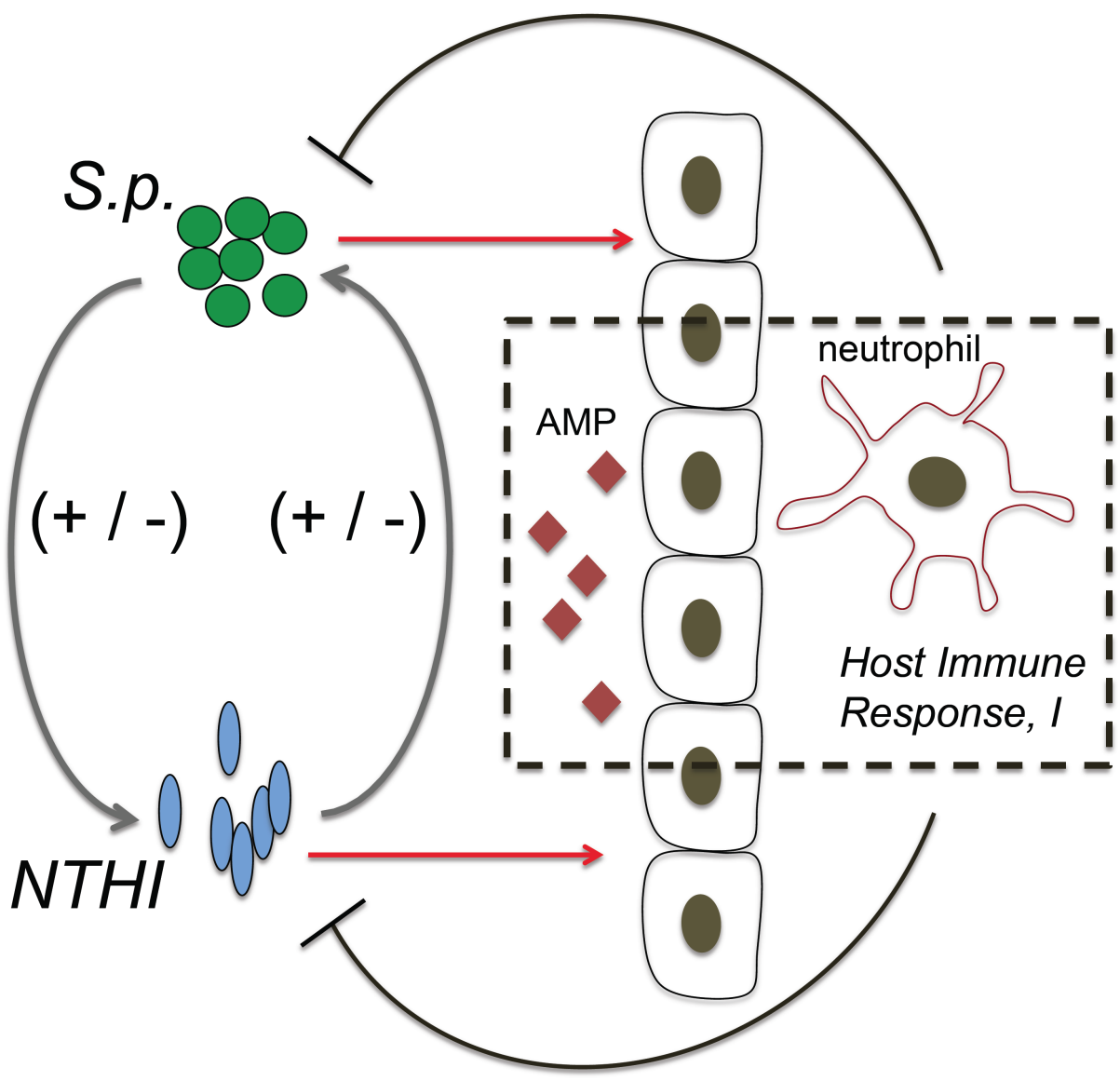

**Fig. S1: Schematic diagram showing the ecological interactions between bacterial species and the host immune system.** NTHI and Sp trigger host innate immune response composed of release of antimicrobial proteins (AMPs) by the epithelial cells and influx of neutrophils in the area of infection which kills NTHI and Sp, possibly with different rates. NTHI and Sp can stay neutral, compete, or, co-operate with each other for growth in the middle ear depending on available nutrients (e.g., essential metals), secreted toxins or quorum sensing molecules.

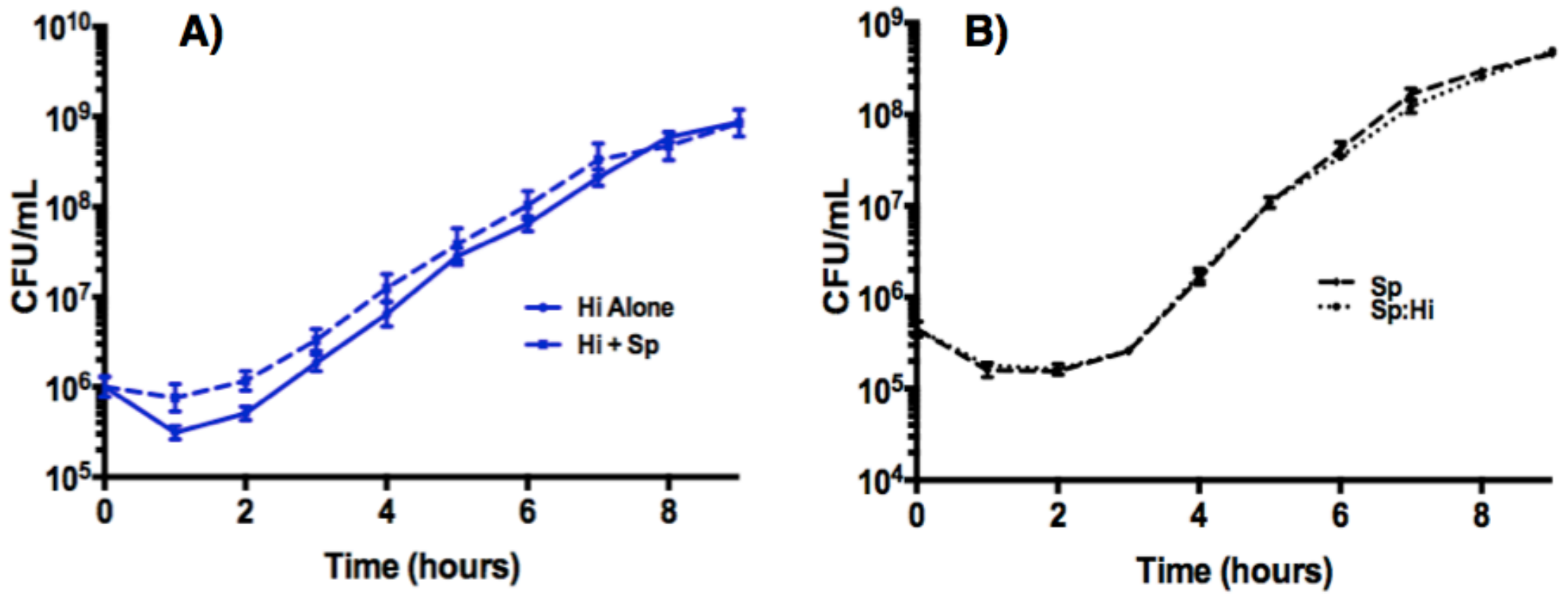

**Fig S2:** A) The growth curve of NTHI alone (solid blue line) and NTHI in the presence of Sp (dashed blue line) as a function of time. The plot is an average of three independent trials. B) The growth curve of Sp alone (dashed black line) and Sp in the presence of NTHI (dotted black line) as a function of time. The plot is an average of three independent trials.

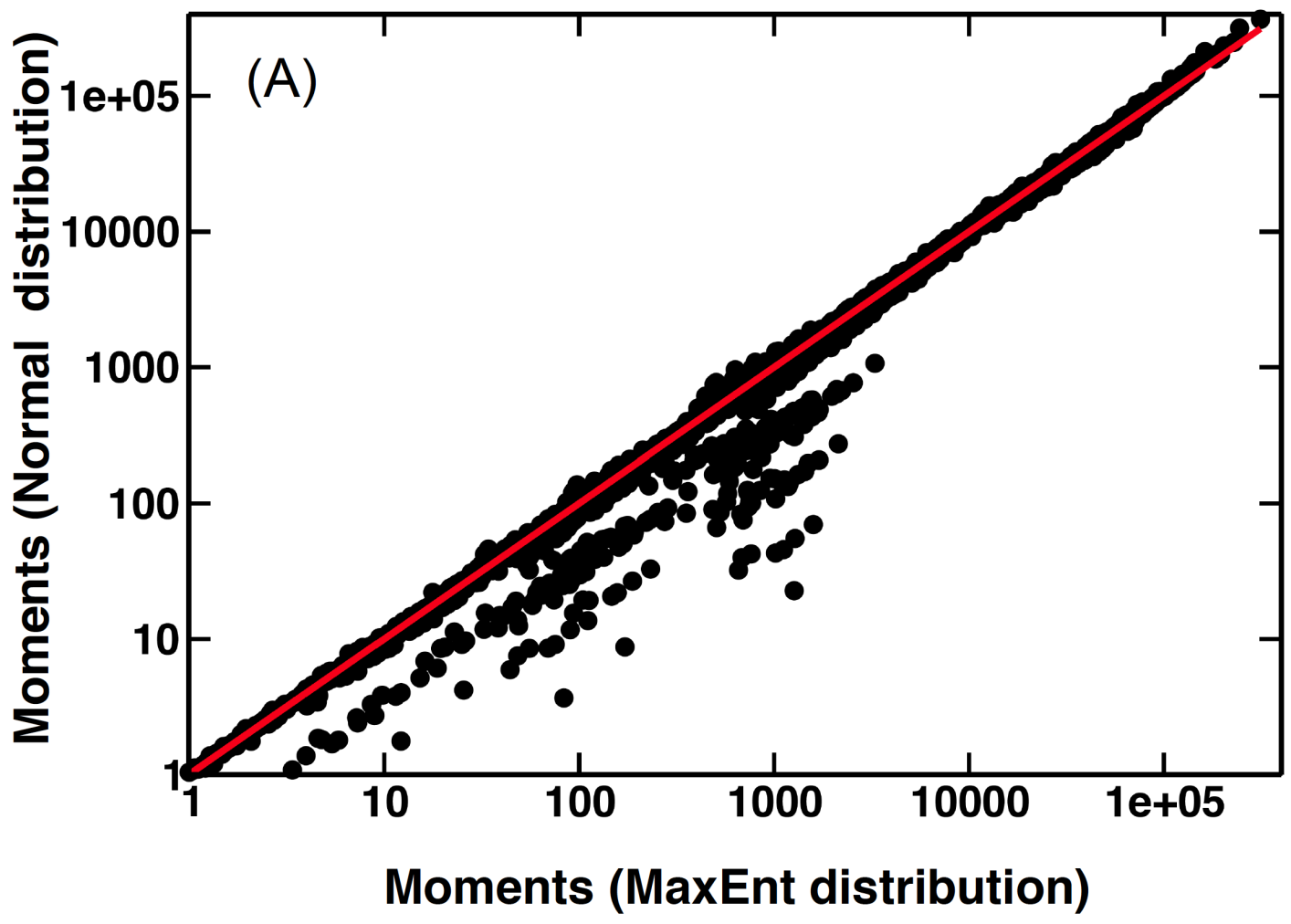

**Fig. S3** $\hat{P}^{(M)}(\{e_i\})$ **can be well approximated by a multivariate normal distribution.** Comparison of 12364 different moments from the $M_{+-}$ model for infection up to the sixth order between the inferred distribution, $\hat{P}^{(M)}(\{e_i\})$, and a multivariate normal distribution which has the same mean values, and second order moments as that of $\hat{P}^{(M)}(\{e_i\})$. About 3% of the total number of moments shown possesses larger values compared to the normal distribution. The majority of the moments lie on the y=x line (red) showing excellent agreement between $\hat{P}^{(M)}(\{e_i\})$ and the normal distribution.